\documentclass[runningheads]{llncs}
\usepackage{graphicx}
\usepackage{amsfonts}
\usepackage{enumitem}
\usepackage{amsmath}
\usepackage{booktabs}
\usepackage{multirow}
\usepackage{enumitem}
\usepackage{subcaption}
\usepackage{color}
\usepackage{url}
\usepackage[ruled,linesnumbered]{algorithm2e}
\newcommand{\eat}[1]{}

\begin{document}
\title{VizPut: Insight-Aware Imputation of Incomplete Data for Visualization Recommendation}
\titlerunning{VizPut: Insight-Aware Imputation of Incomplete Data for VizRec}
%
\author{Rischan Mafrur\inst{1}\orcidID{0000-0003-4424-3736} \and
Mohamed A. Sharaf\inst{2}\orcidID{0000-0002-3405-5224} \and
Guido Zuccon\inst{1}\orcidID{0000-0003-0271-5563}}
\authorrunning{R. Mafrur et al.}
%
\institute{
The University of Queensland, Brisbane, Australia\\
\email{\{r.mafrur,g.zuccon\}@uq.edu.au}\\
 \and
United Arab Emirates University, Al Ain, UAE\\
\email{msharaf@uaeu.ac.ae}}
\maketitle              
\vspace{-15pt}
\begin{abstract}
In insight recommendation systems, obtaining timely and high-quality recommended visual analytics over incomplete data is challenging due to the difficulties in cleaning and processing such data. Failing to address data incompleteness results in diminished recommendation quality, compelling users to impute the incomplete data to a cleaned version through a costly imputation strategy. 
This paper introduces VizPut scheme, an insight-aware selective imputation technique capable of determining which missing values should be imputed in incomplete data to optimize the effectiveness of recommended visualizations within a specified imputation budget.
The VizPut scheme determines the optimal allocation of imputation operations with the objective of achieving maximal effectiveness in recommended visual analytics. We evaluate this approach using real-world datasets, and our experimental results demonstrate that VizPut effectively maximizes the efficacy of recommended visualizations within the user-defined imputation budget.

\eat{
Firstly, we introduce \textit{PriorityBasedOnCellImpact (PBCI)}, a novel insight-aware approach for prioritizing missing values imputation based on the cell contribution (i.e., impact) to the recommendation results, following an "impute-first-insight-next" strategy, where the data is first imputed and then top-k insights are generated. However, this approach may incur an analytics bottleneck, as the cost of data cleaning is often higher than the cost of generating insights, requiring analysts to wait for the completion of data cleaning before generating insights.
To address this challenge, we propose \textit{PriorityBasedOnTemporaryRanking (PBRank)}. The approach follows a "insight-first-impute-next" strategy, where temporary top-k visual insights (i.e., \textit{temp-rank}) are generated from the incomplete data first and the imputation is performed based on the \textit{temp-rank}. This approach focuses the imputation operations only on the data relevant to the candidate top-k insights. 
In this paper, we also propose a hybrid approach that combines PBCI and PBRank. This approach provides two benefits: (1) priority imputation based on cell impact and (2) focused imputation operations on the candidate top-k insights.
}
\end{abstract}
%
%
%
\section{Introduction}
\vspace{-10pt}

\label{sec:intro}
The rapid growth of data in various domains has led to an increasing demand for effective data analysis and visualization tools. Tools such as Tableau~\cite{Tableau}, Spotfire~\cite{Spotfire}, and Power BI~\cite{PowerBI} have been introduced to provide visually appealing visualizations that reveal meaningful insights. However, selecting combination of dimensional attributes and measure attributes that lead to meaningful visualizations without prior knowledge of the data can be a challenging task for an analyst. Manually searching for insights in each visualization is time-consuming and labor-intensive.

This challenge has motivated research efforts to automatically recommend important visualizations based on metrics that capture their utility (e.g., \cite{DiVE}, \cite{VizDeck}, \cite{Profiler}, \cite{SeeDB2015}, \cite{MuVE2016}, \cite{ExtractTopK}, \cite{DeepEye}, \cite{QuickInsight}, \cite{Viewseeker}, \cite{VisGNN}, \cite{SpotLight}, \cite{VizGRank}). These visualization recommendation systems have emerged as a powerful solution to assist users in exploring and understanding complex datasets by automatically recommending the most interesting or important visualizations. However, the performance of these systems is heavily dependent on the quality of the underlying data. Real-world datasets are often fraught with issues, such as noise, inconsistencies, and incompleteness, which can adversely affect the quality of the recommended visualizations (e.g., \cite{taxonomy_dirty_data}, \cite{Profiler}).

In this paper, we focus on addressing the challenges posed by incomplete data in visualization recommendation systems. Incomplete data is a pervasive problem in real-world datasets, as data can be missing for various reasons, such as system failures, human errors, or unavailability of information. Existing visualization recommendation systems typically assume that the analyzed data is clean and complete, which is often not the case in practice. Consequently, there is a pressing need for robust methods to handle incomplete data and improve the overall effectiveness of these systems.

A variety of imputation techniques exist for addressing incomplete data, encompassing crowd-sourcing platforms like Amazon Mechanical Turk (AMT)\footnote{\url{https://www.mturk.com/}} and CrowdFlower\footnote{\url{https://visit.figure-eight.com/People-Powered-Data-Enrichment_T}}. These crowd-sourced approaches have been utilized to achieve data completion or rectification tasks, resulting in high-quality output but requiring substantial human effort. Rule-based cleaning represents an intermediate solution, wherein human expertise is consulted to formulate cleaning rules while automating the repair process. An exemplary case is the study conducted in (e.g., \cite{nadeef}, \cite{VisClean}), which demonstrates a user interface for editing cleaning rules and performing automated cleaning operations. However, in the context of large datasets, data imputation may prove to be costly and ineffective. As an alternative, works such as ImputeDB \cite{DynamicImputation} can be employed to address the challenges posed by incomplete data in the context of large datasets.

ImputeDB's core principle posits that imputation requires execution exclusively on data relevant to a particular query, where this subset is often significantly smaller than the complete database. ImputeDB departs from conventional imputation techniques (e.g., \cite{statistical_imputation}, \cite{mean_knn_imputation}, \cite{clustering_imputation}) which generally operate over the entire dataset. ImputeDB redirects attention from the imputation algorithms towards the identification of optimal imputation operations that improve query results, which lies in devising optimization algorithms offering Pareto-optimal trade-offs between imputation cost and result quality.

In this paper, we share a similar focus with ImputeDB and introduce VizPut, a scheme that determines which missing cells should be prioritized for imputation to maximize the efficacy of recommended visualizations. This strategy can be incorporated with existing crowd-sourcing platforms, enabling analysts to define their preferred imputation budget. VizPut optimally allocates the imputation budget to preserve the high quality of recommended visualizations while adhering to the specified budget constraints.

To illustrate the importance of VizPut, consider the scenario of a data analyst using a crowd-sourcing service or hiring expert to clean their dataset. In many cases, the budget allocated for data cleaning may be insufficient to cover the entire process. For instance, the total cost of cleaning all missing cells might be \$10,000, while the data analyst only has a budget of \$1,000. VizPut enables the data analyst to identify which missing values should be imputed first to optimize the recommendation results within the constraints of their budget, ensuring that the high quality insights are still accessible despite limited resources.

In addressing this challenge, we introduce three types of VizPut scheme for selecting missing cells to be imputed, encompassing \textit{Cell-aware VizPut}, \textit{Ranking-aware VizPut}, and a \textit{Hybrid} that merges the strengths of both Cell-aware VizPut and Ranking-aware VizPut. These methods accommodate various user preferences and scheme contexts, providing a versatile and all-encompassing solution to the difficulties arising from incomplete data in visualization recommendation systems.

The process of performing data imputation in an insight recommendation system typically involves executing data cleaning prior to generating insights. In this context, we propose \textit{Cell-aware VizPut} approach, which is a heuristic method that prioritizes the imputation of missing cells based on their impact on recommendation results (i.e., the number of visualizations affected when the missing cell is imputed). Consequently, this approach identifies which missing cells should be imputed first, and these cells are imputed before generating the top-$k$ visual insights. In this case, we adhere to the traditional approach grounded in the principle of "impute-first-insight-next."
Nonetheless, as previously discussed, our focus aligns with ImputeDB, which posits that imputation should only be performed on relevant data. In this regard, we propose \textit{Ranking-aware VizPut}, a prioritization approach premised on the temporary ranking of visualizations (i.e., \textit{temp-rank}). Incomplete data is initially utilized to generate recommended visualizations, followed by the imputation process based on the \textit{temp-rank} until the imputation budget is exhausted. The underlying concept of this approach is that the imputation budget should be prioritized for candidate top-$k$ insights, necessitating the generation of \textit{temp-rank} from the incomplete data first. However, it is important to note that the \textit{temp-rank} may be misleading, as it is derived from incomplete data. To address this issue, we extend \textit{Ranking-aware VizPut} with an alternative weighting scheme.
Finally, we also propose Hybrid approach that combines both the Cell-aware VizPut and Ranking-aware VizPut. The Hybrid approach provides the benefits of both Cell-aware VizPut and Ranking-aware VizPut methods where the imputation budget is optimized towards the candidate top-k insights (Ranking-aware VizPut), and the selected missing cells are based on their highest contribution score (Cell-aware VizPut). In summary, our contributions are:

\begin{itemize}
    \item We propose the \textit{Cell-aware VizPut}, an insight-aware selective imputation technique that prioritizes the imputation of missing cells based on their impact on recommendation results (Sec.~\ref{subsubsection:VizPut-Cell}).
    \item The \textit{Ranking-aware VizPut} is proposed as an insight-aware imputation strategy based on temporary ranking that selectively places imputation operations only on data relevant to the top-k candidate insights (Sec.~\ref{subsection:VizPut-Ranking}).
    \item We introduce a hybrid approach that combines the strengths of \textit{Cell-aware VizPut} and \textit{Ranking-aware VizPut} (Sec.~\ref{subsection:hybrid}).
    \item An extensive experimental evaluation on real datasets is conducted to compare the performance of our proposed approaches with baselines (Sec.~\ref{sec:experiment_evaluation_vizput}).
\end{itemize}

%
\vspace{-10pt}
\section{Preliminaries and Related Work}
\vspace{-10pt}

\label{sec:recommending_visual_insight}
This section presents an overview for recommending top-k visual insights. Firstly, we describe the methodology for generating such insights (Sec.~\ref{subsection:recommending_visual_insight}). Subsequently, we outline the challenges faced in generating top-k visual insights from incomplete data and provide a formal problem definition for generating top-k visual insights from incomplete data (Sec.~\ref{subsection:incomplete_data_problem}).

\vspace{-8pt}
\subsection{Recommending Visual Insights}
\label{subsection:recommending_visual_insight} 
\vspace{-5pt}

To recommend visual insights, we consider a complete multi-dimensional dataset $D$. The dataset $D$ is comprised of a set of dimensional attributes $\mathbb{A}$ and a set of measure attributes $\mathbb{M}$.
Also, let $\mathbb{F}$ be a set of possible aggregate functions over measure attributes such as {\ttfamily COUNT, AVG, SUM, MIN and MAX}. 
Hence, specifying different combinations of dimension and measure attributes along with various aggregate functions, generates a set of possible visualizations $\mathbb{V}$ over $D$.
For instance, a possible visualization $V_i$ is specified by a tuple $<\!A_i$, $ M_i$, $F_i\!>$, where $A_i \in \mathbb{A}$, $M_i \in \mathbb{M}$, and  $F_i \in \mathbb{F}$, and it can be formally defined as:
$Vi$ {{\ttfamily : VISUALIZE bar (SELECT A, F(M) FROM D WHERE T GROUP BY A)}}. Where {{\ttfamily VISUALIZE}} specifies the visualization type (i.e., bar chart), {{\ttfamily SELECT}} extracts the selected columns which can be dimensional attributes $A \in \mathbb{A}$ or measures $M \in \mathbb{M}$, $T$ is the query predicate (e.g., disease = ’Yes’), and {{\ttfamily GROUP BY}} is used in collaboration with the {{\ttfamily SELECT}} statement to arrange identical data into groups. A visualization $V_i$ is only possible to obtain if the analyst knows exactly the parameters (e.g., dimensional attributes, measures, aggregate functions, grouping attributes, etc.), which specify some aggregate visualizations that lead to valuable visual insights. This iterative process of creating and refining visualizations to uncover valuable insights can be time-consuming.
Several solutions for recommending visualizations have recently emerged to address the need for efficient data analysis and exploration (e.g., \cite{SeeDB2015}, \cite{TKDEHumaira}, \cite{DiVE}, \cite{ExtractTopK}, \cite{DeepEye}, \cite{QuickInsight}, \cite{Foresight}, \cite{SpotLight}, \cite{VisGNN}, \cite{VizGRank}). These solutions generate a large number of possible visualizations $\mathbb{V}$ and rank them based on metrics (e.g., deviation-based approach) that capture the \textit{utility} of the recommendations. Finally, top-k visual insights are recommended to users. 

Previous studies (e.g., \cite{SeeDB2015}, \cite{SeeDB2014}) have demonstrated the effectiveness of the deviation-based approach in presenting interesting visualizations that reveal distinctive trends of analyzed datasets. The deviation-based approach compares an aggregate visualization generated from the selected subset dataset $D_Q$ (i.e., target visualization $V_i(D_Q)$) to the same visualization if generated from a reference dataset $D_R$ (i.e., reference visualization $V_i(D_R)$). To calculate the outstanding/deviation score, each target visualization $V_i(D_Q)$ is normalized into a \textit{probability distribution} $P_b[V_i(D_Q)]$ and similarly, each reference visualization into $P_b[V_i(D_R)]$. In particular, consider an aggregate visualization $V_i=<\!A,M,F\!>$. The result of that visualization can be represented as the set of tuples: $<\!(a_1, g_1), (a_j, g_j), ..., (a_t, g_t)\!>$, where $t$ is the number of distinct values (i.e., groups) in attribute $A$, $a_j$ is the $j$-th group in attribute $A$, and $g_j$ is the aggregated value $F(M)$ for the group $a_j$. Hence, $V_i$ is normalized by the sum of aggregate values $G=\sum\limits_{j=1}^{t} g_j$, resulting in the probability distribution $P_b[V_i] = <\!\frac{g_1}{G}, \frac{g_2}{G}, ..., \frac{g_t}{G}\!>$. Finally, the utility score of $V_i$ is measured in terms of the distance between $P_b[V_i(D_Q)]$ and $P_b[V_i(D_R)]$, and is simply defined as: $U\left(V_i\right) = dist\left(P_b\left[V_i(D_Q)\right], P_b\left[V_i(D_R)\right]\right)$.
The process of generating recommended visualizations can be summarized as three layers:

\begin{itemize}[noitemsep,topsep=0pt]
\item \textit{Generating Visualizations}: All possible visualizations are generated from the selected subset dataset $D_Q$ (i.e., target visualization $V_i(D_Q)$) and a reference dataset $D_R$ (i.e., reference visualization $V_i(D_R)$).
\item \textit{Calculating Importance Score}: The deviation-based approach is used to compare the aggregate visualization generated from $D_Q$ (i.e., $V_i(D_Q)$) to the same visualization if generated from $D_R$ (i.e., $V_i(D_R)$).
\item \textit{Ranking and Presenting Recommendations}: Visualizations are ranked based on their importance score, and the top-k visualizations are recommended to the user.
\end{itemize}

\subsection{Handling Incomplete Data in Visualization Recommendation}
\label{subsection:incomplete_data_problem}

\begin{figure}[t!]
\centering
\includegraphics[width=4.5in]{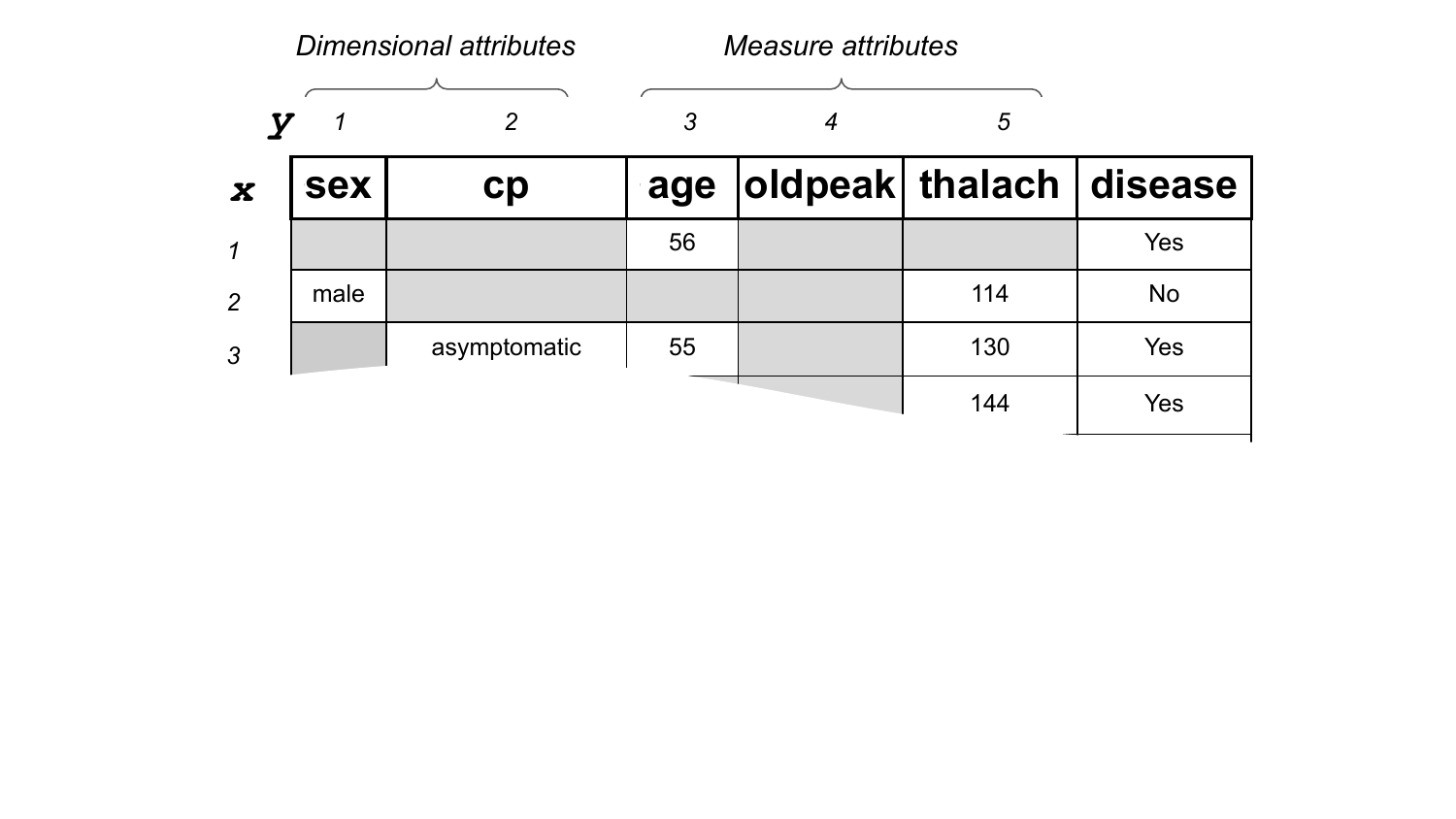}
\vspace{-80pt}
\caption{An example of incomplete data}
\label{fig:incomplete_data_example_empty}
\end{figure}

Handling incomplete data in visualization recommendation systems is a challenging task. Our prior work in \cite{QualityMatters} demonstrates the impact of incomplete data on visualization recommendation results. A user analyzing data with 20\% missing values will obtain significantly different top-k recommended visualizations compared to those obtained from a complete dataset, resulting in incorrect insights. Given the prevalence of incomplete data, it is crucial to develop methods for addressing this challenge. 

Consider the incomplete data in Figure \ref{fig:incomplete_data_example_empty}, which consists of two dimensional attributes $\mathbb{A}$ and three measure attributes $\mathbb{M}$. 
In the figure, grey cells indicate incomplete cells, while white cells indicate complete cells. 
To produce recommended visual analytics of high quality, the imputation must be performed on the incomplete data before visualizations are generated.
As mentioned in Sec.~\ref{sec:intro}, data imputation is expensive! Suppose a data analyst has a limited imputation budget of five cells. The challenge is to select which five cells from the missing cells in Figure \ref{fig:incomplete_data_example_empty} to be imputed in order to maximize the effectiveness of recommended visualizations. 

In this study, it is important to emphasize that we are not proposing a new data imputation technique. Numerous imputation methods have been developed, such as Mean Imputation and K-Nearest Neighbor imputation (\textit{e.g.}, \cite{mean_knn_imputation}, \cite{statistical_imputation}). One example is Clustering-Based Imputation \cite{clustering_imputation}, which estimates missing values using the nearest neighbor within the same cluster constructed based on non-missing values. More recently, advanced imputation models have adopted machine learning approaches (\textit{e.g.}, \cite{ML_partial_complete_data}, \cite{ML_imputation}). Generally, these existing imputation methods operate over the entire dataset, replacing all missing values with predicted values. However, executing a sophisticated imputation algorithm on a large dataset can be computationally demanding. For instance, \cite{DynamicImputation} illustrates that even a relatively simple decision tree algorithm requires nearly 6 hours to train and run on a database with 600,000 rows. As an alternative to this challenge, sampling techniques can also be employed in data imputation (\textit{e.g.}, \cite{SampleClean}, \cite{AQP}), allowing for the imputation of only a sample of the data without necessitating imputation on the entire dataset.

In contrast to prior research, our study emphasizes not on the imputation techniques themselves, as nearly any such method can be utilized, our emphasis is on optimally positioning imputation operations within the incomplete data, with the aim of maximizing the efficacy of the recommended visualizations. In pursuit of our objective, this paper employs the original complete dataset as the ground truth data. We create a duplicate of the complete data, introduce missing values with varying distribution patterns, and refer to this as incomplete data. Subsequently, we employ VizPut to impute the missing values in the incomplete data in accordance with a predetermined imputation budget. It is important to note that VizPut primarily focuses on prioritizing the missing cells, with the imputed values sourced from the ground truth data, thereby ensuring accuracy. Our central aim is to identify the most effective methods concerning which missing cells should be prioritized in order to achieve optimal effectiveness of recommendation results

Consider a set of top-k visualizations, denoted as $S_c$, generated from a multi-dimensional dataset $D$. Suppose $D_I$ represents an incomplete version of $D$. To generate recommended visual analytics, we must impute $D_I$, given an imputation budget $g$ and the imputed version of $D_I$ denoted by $D_f$. For the sake of discussion, let $S_c$ and $S_f$ be the sets of top-k visualizations from the complete and imputed data, respectively.
To evaluate the priority functions we propose, we compare the recommended visual analytics $S_f$ derived from the imputed data $D_f$ with the top-k set obtained from the complete data $S_c$. We employ various metrics from our prior work \cite{QualityMatters} to gauge the quality of the recommended visualizations in $S_f$ relative to $S_c$. First, we apply the {\em Jaccard distance} \cite{Jaccard}, which assesses the composition of two sets. The Jaccard distance score is determined by dividing the number of common visualizations by the total number of visualizations. Consequently, when applied to set comparison, sets with identical compositions yield identical similarity scores.
Nonetheless, our work considers the order of visualizations in the top-k set to be crucial. For example, the top-1 visualization is more significant than the top-10 visualization. Therefore, we employ the second metric, {\em Rank Biased Overlap (RBO)} \cite{RBO}, to account for visualization ranking while evaluating recommendation quality. RBO takes into consideration both the composition and ranking of the two sets.

Our objective is to identify way to find the missing values that need to be imputed first such as that yields recommended visual analytics $S_f$ generated from the imputed data $D_f$, closely approximating $S_c$ derived from the complete data $D$. We can formally define this problem as follows:

\begin{definition} 
\label{def_vizput}
\textbf{Recommending Top-k Visual Insights from Incomplete Data.} Given a set of top-k visualizations $S_c$ generated from the complete data $D$, let $D_I$ represent an incomplete version of $D$ and $g$ denote the imputation budget. The objective is to select and impute $g$ missing cells to create an imputed data version $D_f$ of $D_I$, such that the effectiveness of the recommended visualizations generated from $D_f$ closely approximates the effectiveness of $S_c$.
\end{definition} 

In order to address the problem defined in Definition~\ref{def_vizput}, we propose VizPut, which comprises various variants of priority functions, described in the following subsections. It is important to note that the symbols used in this paper are summarized in Table~\ref{tab:tab-symbols}.

\begin{table}
	\centering
	\caption{Table of symbols}
	\label{tab:tab-symbols}
	\begin{tabular}{lll}
		\toprule
		Symbols &Description\\
		\midrule
		$k$ & the size of top-k recommended views\\
		$S$ & set of views\\
		$ \mathbb{V}  $& set of all possible views\\
		$ A $ & a dimensional attribute \\
		$ M  $& a measure attribute\\ 
		$ F $ & aggregate function \\
		$ Q $ & a user query \\
		$ D  $& a multi-dimensional database \\
		$ D_Q  $& a target subset of $ D_B  $\\
		$ D_R  $& a reference subset of $ D_B  $\\
        $ D_I  $& an incomplete data version of $ D_B  $\\
        $ g  $& imputation budget\\
        $ D_f  $& an imputed data version of $ D_I  $\\
		$ V_i  $& a view query\\
        $C_{x,y}$ & a cell with $x$ represents row and $y$ represents column \\
		$P(C_{x,y})$ & priority function can be a single function or a combination of two or more functions \\
		$ N(C_{x,y})$  & contribution score of a missing cell $C_{x,y}$ \\
		$ F(C_{x,y})$ &  fairness score of a column that mapped to a missing cell $C_{x,y}$\\
        $ R(C_{x,y})$ &  ranking score of visualization that mapped to a missing cell $C_{x,y}$\\
        $ \hat{R}(C_{x,y})$ &  ranking and weighting score of visualization that mapped to a missing cell $C_{x,y}$\\
		\bottomrule
	\end{tabular}
\end{table}

\vspace{-10pt}
\section{VizPut: Insight-Aware Imputation of Incomplete Data for Visualization Recommendation}
\vspace{-10pt}

In this section, we first discuss our baseline solutions
for missing cells selection (Sec.~\ref{subsubsection:baseline_vizput}). Then, we present our proposed approach Cell-aware VizPut (Sec.~\ref{subsubsection:VizPut-Cell}) and Ranking-aware VizPut (Sec.~\ref{subsection:VizPut-Ranking}).


\subsection{Baseline Solutions}
\label{subsubsection:baseline_vizput}
Consider the incomplete data in Figure \ref{fig:incomplete_data_example_empty}, suppose a data analyst has a limited imputation budget of five cells. The challenge is to select which five cells from the missing cells in Figure \ref{fig:incomplete_data_example_empty} to be imputed. We define three baseline methods as follows:
\begin{itemize}[noitemsep,topsep=0pt]
\item \textit{No Imputation}: In this approach, recommended visualizations are generated from incomplete data without any imputation. The cost of imputation is zero since the budget is not utilized, but the effectiveness of recommended visualizations may be low due to missing values.
\item \textit{Random Selection Imputation}: In this approach, five missing cells are randomly selected for imputation from the incomplete data. The effectiveness of recommended visualizations may be better compared to No Imputation as the number of missing cells is lower, but the imputation cost will still be higher than No Imputation.
\item \textit{Fairness Imputation:} This approach involves selecting five missing cells based on a higher ratio of missing cells in a column. To elaborate, prior to choosing a missing cell, a fairness score is computed for each column. This score can be determined by calculating the ratio of the number of missing cells to the total number of cells within a given column. In every iteration, a missing cell is selected from the column exhibiting the highest fairness score.


\end{itemize}

\label{sec:insight_aware_approach}

\begin{figure}[t!]
\centering
\includegraphics[width=5in]{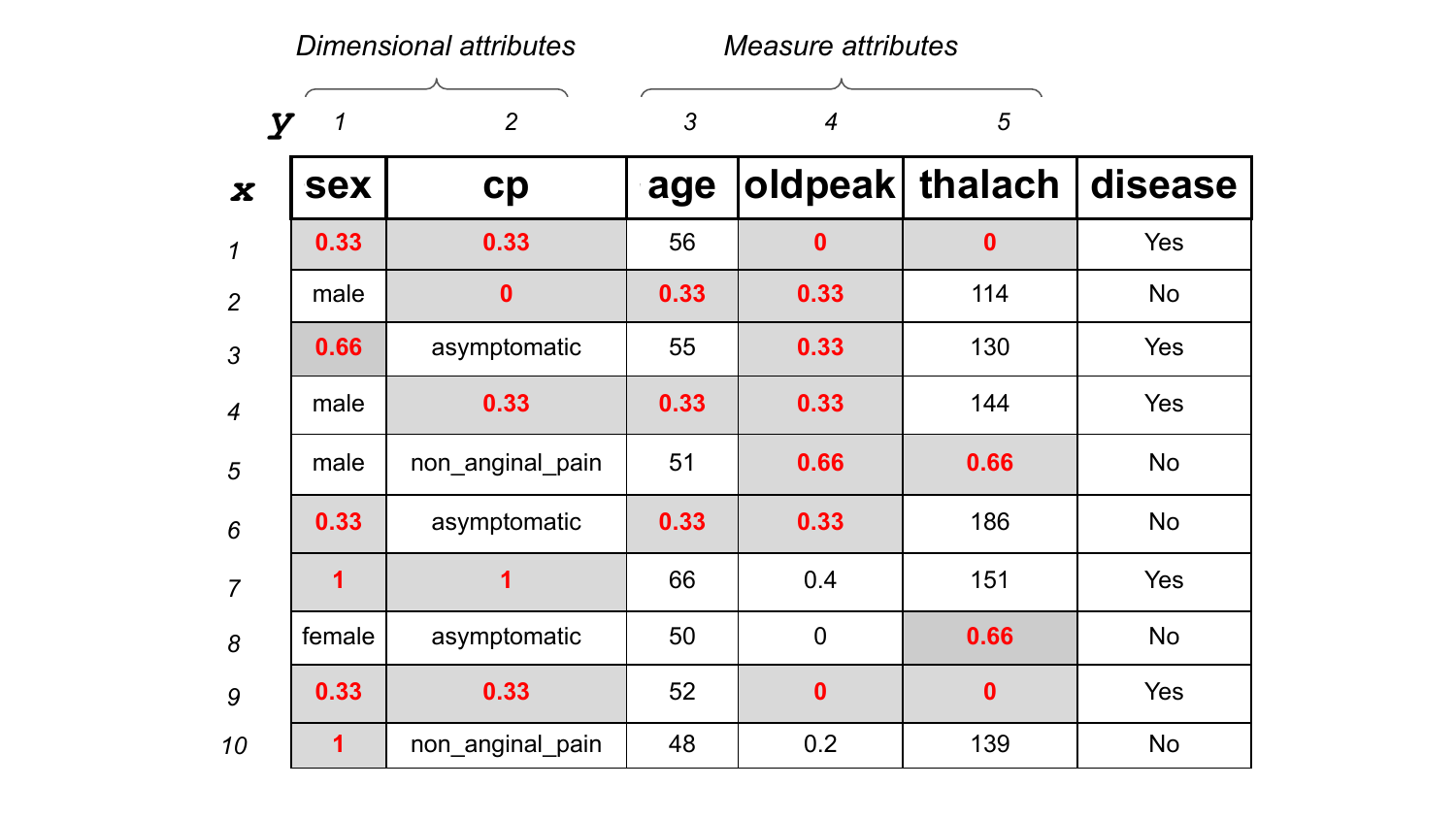}
\caption{The example of incomplete data}
\label{fig:incomplete_data_example}
\end{figure}

\setlength{\textfloatsep}{0pt}
\begin{algorithm}[t]
\DontPrintSemicolon
\SetKwBlock{Begin}{}{end}
	\KwIn{Incomplete data $D_I$, imputation budget $g$ }
	\KwOut{Imputed data $D_f$}  
	 \SetKwFunction{FMain}{VizPutBasedOnCell}
     $\mathbb{A} \leftarrow $ get all dimensional attributes from $D_I$\; 
    $\mathbb{M} \leftarrow $ get all measures attributes from $D_I$\; 
    $ C_{x,y}  \leftarrow $ a missing cell(row $i$, column $j$) on $ D_I $\;   
    $ D_f  \leftarrow D_I $\;  
    $ N  \leftarrow 0 $\;  
    \SetKwProg{Fn}{Function}{:}{}
    \Fn{\FMain{$D_I$}}{
        {\eIf{$ C_{x,y} $ in $\mathbb{A}$ }
            {\textit{get the number of corresponding visualizations} $(N)$\;
            $ N \leftarrow $ count complete cells in $\mathbb{M}_x$;}
            {$ N \leftarrow $ count complete cells in $\mathbb{A}_x$;}

        }
        \textbf{return} $ C_{x,y} $, $ N $ 
    }
    \textbf{End Function}
	
	\While{g != 0}{
		$ C_{x,y} \leftarrow argmax (VizPutBasedOnCell(D_I))$\;
		$ D_f.impute\left(C_{x,y}\right)$\;
              $ D_I \leftarrow D_f$\;
		$ g$ $ -= 1$\;
	}
	return $D_f$
	\caption{VizPut-Cell algorithm}
	\label{VizPut_Cell_Impact_algorithm}
\end{algorithm}

\subsection{Cell-aware VizPut}
\label{subsubsection:VizPut-Cell}
The concept behind \textit{Cell-aware VizPut} is selecting missing cells based on their maximum contribution to the recommendation results. In this work, we propose some variants of Cell-aware VizPut, including \textit{VizPut-Cell}, \textit{VizPut-Cell(f)}, \textit{VizPut-Cell(f, v)}, which will be explained subsequently.

\subsubsection{VizPut-Cell}

The VizPut-Cell is a heuristic priority function for selecting missing cells based on their maximum contribution to the recommendation results. The contribution of a cell is quantified by counting the number of visualizations associated with it.

Consider $C_{x,y}$, which represents a cell with $x$ as the row number and $y$ as the column number, where $x = {1,2,..,10}$ and $y= {1,2,..,5}$. 
Let $\hat{a}$ represents the number of complete cells in $\mathbb{A}x$ and $\hat{m}$ represents the number of complete cells in $\mathbb{M}x$.
The contribution score of $C_{x,y}$, denoted as $N$, is calculated as the normalized of number of corresponding visualizations $V_i$ to $C_{x,y}$, which is formally defined as: 

\begin{equation}
N(C_{x,y}) = \frac{\hat{m}}{max(\hat{a}, \hat{m})},\hspace{5pt} where \hspace{5pt} C_{x,y} \in \mathbb{A} \quad | \quad
N(C_{x,y}) = \frac{\hat{a}}{max(\hat{a}, \hat{m})},\hspace{5pt} where \hspace{5pt} C_{x,y} \in \mathbb{M}
\end{equation}

The priority function $P(C_{x,y})$ in VizPut-Cell is defined as:

\begin{equation}
P(C_{x,y}) = N(C_{x,y})
\label{eq:cell_impact_final}
\end{equation}

In this equation, $P(C_{x,y})$ represents the priority score of cell $C_{x,y}$, while $N(C_{x,y})$ denotes its contribution score to the recommendation results.

For instance, as depicted in Figure \ref{fig:incomplete_data_example}, the columns $y$ comprise two-dimensional attributes $\mathbb{A}$ (i.e., \textit{sex, cp}) and three measure attributes $\mathbb{M}$ (i.e., \textit{age, oldpeak, thalach}). Let us consider $C_{1,1}$, with $(x=1, y=1)$, as an example. The priority score of $P(C_{1,1})$ equals $0.33$, as the cell $C_{1,1}$ possesses a single contribution owing to its association with only one complete cell in $\mathbb{M}$, specifically, $age=56$. Imputing $C_{1,1}$ would impact a single visualization, $V_1 = <sex, F(age)>$. The priority score $0.33$ originates from the normalized contribution score, $\frac{1}{3}$, where $1$ denotes the contribution score $\hat{m}$ and $3$ represents the maximum contribution scores $max(\hat{a}, \hat{m})$ from the incomplete data $D_I$.
Conversely, cells with the highest contribution (i.e., a score of $1$), such as $C_{7,1}$, $C_{7,2}$, and $C_{10,1}$, have three contribution scores each, as imputing them can influence multiple visualizations. For instance, $C_{7,1}$ contributes to $V_1 = <sex, F(age)>$, $V_2 = <sex, F(oldpeak)>$, and $V_3 = <sex, F(thalach)>$; $C_{7,2}$ affects $V_1 = <cp, F(age)>$, $V_2 = <cp, F(oldpeak)>$, and $V_3 = <cp, F(thalach)>$; and $C_{10,1}$ contributes to $V_1 = <sex, F(age)>$, $V_2 = <sex, F(oldpeak)>$, and $V_3 = <sex, F(thalach)>$ (Algorithm~\ref{VizPut_Cell_Impact_algorithm} line 9 and 11).

Assuming the data analyst has a limited budget for missing cell imputation (i.e., $g=5$), missing cells with the highest priority score of $1$, such as $C_{7,1}$, $C_{7,2}$, and $C_{10,1}$, will be chosen. The subsequent candidates for missing cell selection have a score of $0.66$, encompassing $C_{3,1}$, $C_{5,4}$, $C_{5,5}$, and $C_{8,5}$. Given that the budget allows for only five cells, just two cells with a $0.66$ score can be selected. In this scenario, the selection is based on random.


A limitation of VizPut-Cell arises when the dataset exhibits an extreme imbalance between the sizes of $|\mathbb{A}|$ and $|\mathbb{M}|$. This imbalance may result in certain attributes being less imputed or not imputed at all compared to others. For instance, consider incomplete data containing a single dimensional attribute $A$ and numerous measure attributes $\mathbb{M}$. Even with an equal distribution of missing values across all columns, Algorithm~\ref{VizPut_Cell_Impact_algorithm} assigns a higher priority to the dimensional attribute due to the nature of the view generated from the combination of $A$ and $\mathbb{M}$. Consequently, given the limited imputation budget, one or more attributes may be less imputed or not imputed at all.
To tackle this challenge, we augment the priority score $P$ of VizPut-Cell by incorporating a fairness parameter. This parameter bears similarity to the Fairness Imputation employed in the baselines, which is detailed in the subsequent section.

\subsubsection{Fairness Awareness in VizPut-Cell}
\label{subsection:VizPut-Cell_fairness_algorithm}

Recall that the limitation of the \textit{VizPut-Cell} method is that it may result in one attribute being either never imputed or less imputed compared to others, especially when $|\mathbb{A}|$ and $\mathbb{|M|}$ are unbalanced. To address this issue, we extend the \textit{VizPut-Cell} with a fairness parameter, resulting in the \textit{VizPut-Cell(f)} approach. This Fairness parameter bears similarity to the Fairness Imputation employed in the baselines. The fairness parameter aims to distribute the imputation budget evenly among all columns in $D_I$. The fairness score is calculated based on the ratio of missing cells to total cells in a column. Thus, a column with a higher ratio of missing cells will have a higher priority score compared to a column with a lower ratio of missing cells.

We define $F$ as the fairness parameter, which represents the normalized of ratio of the quantity of missing cells to the overall number of cells in a column. Let us denote $C_y$ as the quantity of missing cells within column $y$ and $T_y$ as the total cell count within the same column. Consequently, we can express $F$ in the following condensed form:

\begin{equation}
F(C_{x,y}) = \frac{\frac{C_y}{T_y}}{max(\frac{C_y}{T_y})}
\end{equation}

To illustrate, consider Figure \ref{fig:incomplete_data_example}, wherein the incomplete data comprises five columns, $y = {1,2,3,4,5}$, each with respective $\frac{C_y}{T_y}$ scores $y_1 = \frac{6}{10}$, $y_2 = \frac{5}{10}$, $y_3 = \frac{3}{10}$, $y_4 = \frac{7}{10}$, and $y_5 = \frac{6}{10}$. Here, column $y_4$ \textit{(oldpeak)} possesses the highest $\frac{C_y}{T_y}$ score, attributable to its extensive number of missing cells. Therefore, in the initial imputation iteration, all missing cells within the \textit{oldpeak} column are assigned a fairness score $F(C_{x,4}) = 0.7/max(\frac{C_y}{T_y}) = 1$. This fairness score undergoes modification with each iterative imputation of cells.



To define the priority function of \textit{VizPut-Cell(f)}, we extend the priority function of \textit{VizPut-Cell}. The priority score for a single missing cell $C_{x,y}$ in \textit{VizPut-Cell(f)} is defined as:

\begin{equation}
P(C_{x,y}) = N(C_{x,y}) \times F(C_{x,y})
\label{eq:fairness_final}
\end{equation}

In this equation, $P(C_{x,y})$ represents the priority score of $C_{x,y}$, $N(C_{x,y})$ denotes its contribution score to the recommendation results, and $F(C_{x,y})$ is the fairness score indicating the missing rate in the column of $C_{x,y}$. By incorporating the fairness score, the column with more missing cells will have a higher priority to be imputed first. Note that the priority function $P(C_{x,y})$ can be expressed either as a singular function or a combination of two or more functions (Equation~\ref{eq:fairness_final}), where multiplication is utilized as the combining operator. While addition can also be used in lieu of multiplication, since that both operators produce similar results as the priority scores of the missing cells will ultimately be ranked, and the selection of the missing cells will be based on priority scores in descending order.

\subsubsection{Impact of target and reference views in VizPut-Cell}
\label{subsection:UsedView}
In this section, we present the impact of number used visualizations in the different settings of target and reference subsets. As described in Sec.~\ref{sec:recommending_visual_insight}, recommended visualizations are generated by comparing a view from the target subset $V_i(D_Q)$ to a view from the reference subset $V_i(D_R)$. There are three common scenarios for the target versus reference subset settings, which are:
\begin{enumerate}[noitemsep,topsep=0pt]
    \item Target subset is a subset of reference subset and reference data is the entire dataset, i.e., $D_Q \subset D_R, D_R = D$. For example, in a heart disease dataset, $D_Q:$ \textit{disease = Yes} and $D_R:$ \textit{whole data};
    \item Target and reference subsets are different subsets, and there is no other subset available, i.e., $D_Q \cup D_R = D$. For example, in a heart disease dataset, $D_Q:$ \textit{disease = Yes} and $D_R:$ \textit{disease = No}. As the $disease$ attribute has only two categories, i.e., $[Yes, No]$, there are only two subsets that can be generated based on the $disease$ predicate;
    \item Target and reference subsets are different subsets, and there is another subset available, i.e., $D_Q \cup D_R \neq D$. For example, in a heart disease dataset, $D_Q:$ \textit{cp = typical angina} and $D_R:$ \textit{cp = atypical angina}. The $cp$ attribute has four categories, and hence, other subsets such as $D_R:$ \textit{cp = non anginal pain} and $D_R:$ \textit{cp = asymptomatic} are available.
\end{enumerate}

To handle these three cases, we propose two variants of \textit{VizPut-Cell}, namely \textit{VizPut-Cell(v)} and \textit{VizPut-Cell(f, v)}. The objective of these approaches is to evaluate the impact of the number of used visualizations, $V$, in different target vs. reference subset settings. Consider that $\hat{v}$ is the frequency of a view being used, such as once, twice, or not at all. $V$ reflects the normalized frequency of a view being used, which can be defined as:

\begin{equation}
V(C_{x,y}) = \frac{\hat{v}}{max(\hat{v})}
\end{equation}

For instance, in the first case, cells belonging to the target subset $D_Q$ have $\hat{v} = 2$ since they are used in both the target and reference subsets, as $D_Q \subset D_R$. On the other hand, cells belonging to the reference data $D_R$ that are not in $D_Q$ have $\hat{v} = 1$. In the third case, cells belonging to \textit{cp = non anginal pain} and \textit{cp = asymptomatic} have $\hat{v} = 0$ since they are not used in any computation. Those $\hat{v}$ values are then divided by the $max(\hat{v})$ to obtain $V$.

To incorporate this $V$ parameter, we extend the priority function $P(C_{x,y})$ of \textit{VizPut-Cell} and \textit{VizPut-Cell(f)}. Therefore, the overall priority function of \textit{VizPut-Cell(v)} is defined as:

\begin{equation}
P(C_{x,y}) = N(C_{x,y}) \times V(C_{x,y})
\end{equation}

Meanwhile, the overall priority function of \textit{VizPut-Cell(f, v)} is defined as:

\begin{equation}
P(C_{x,y}) = N(C_{x,y}) \times F(C_{x,y}) \times V(C_{x,y})
\end{equation}

In these equations, $P(C_{x,y})$ represents the overall priority score, $N(C_{x,y})$ is the contribution score, $F(C_{x,y})$ is the fairness score, and $V(C_{x,y})$ is the normalized number of used views.

\subsection{Ranking-aware VizPut}
\label{subsection:VizPut-Ranking}

The fundamental premise of the \textit{Cell-aware VizPut} method is to tackle incomplete data prior to generating insights, aiming to enhance the quality of recommended visualizations. For instance, the selection of missing cells in \textit{VizPut-Cell} is based on the count of visualizations that are affected when the cell is imputed. However, it remains unclear whether the chosen cell is pertinent to the candidate top-$k$ insights. As a result, we introduce an alternative approach, \textit{Ranking-aware VizPut}. In this method, imputation is performed on the candidate top-$k$ insights. To obtain these candidate insights, the top-$k$ visual insights must be generated from the incomplete data \textit{temp-rank}, despite the awareness that such rankings might be misleading. This approach adheres to the insight-first-impute-next strategy, in contrast to \textit{VizPut-Cell}, which follows the impute-first-insight-next approach. \textit{Ranking-aware VizPut} includes two variants: \textit{VizPut-Ranking} and \textit{VizPut-Ranking(w)}, which will be further elucidated subsequently.

\subsubsection{\textit{VizPut-Ranking}}

\begin{figure}[t!]
\centering
\includegraphics[width=5in]{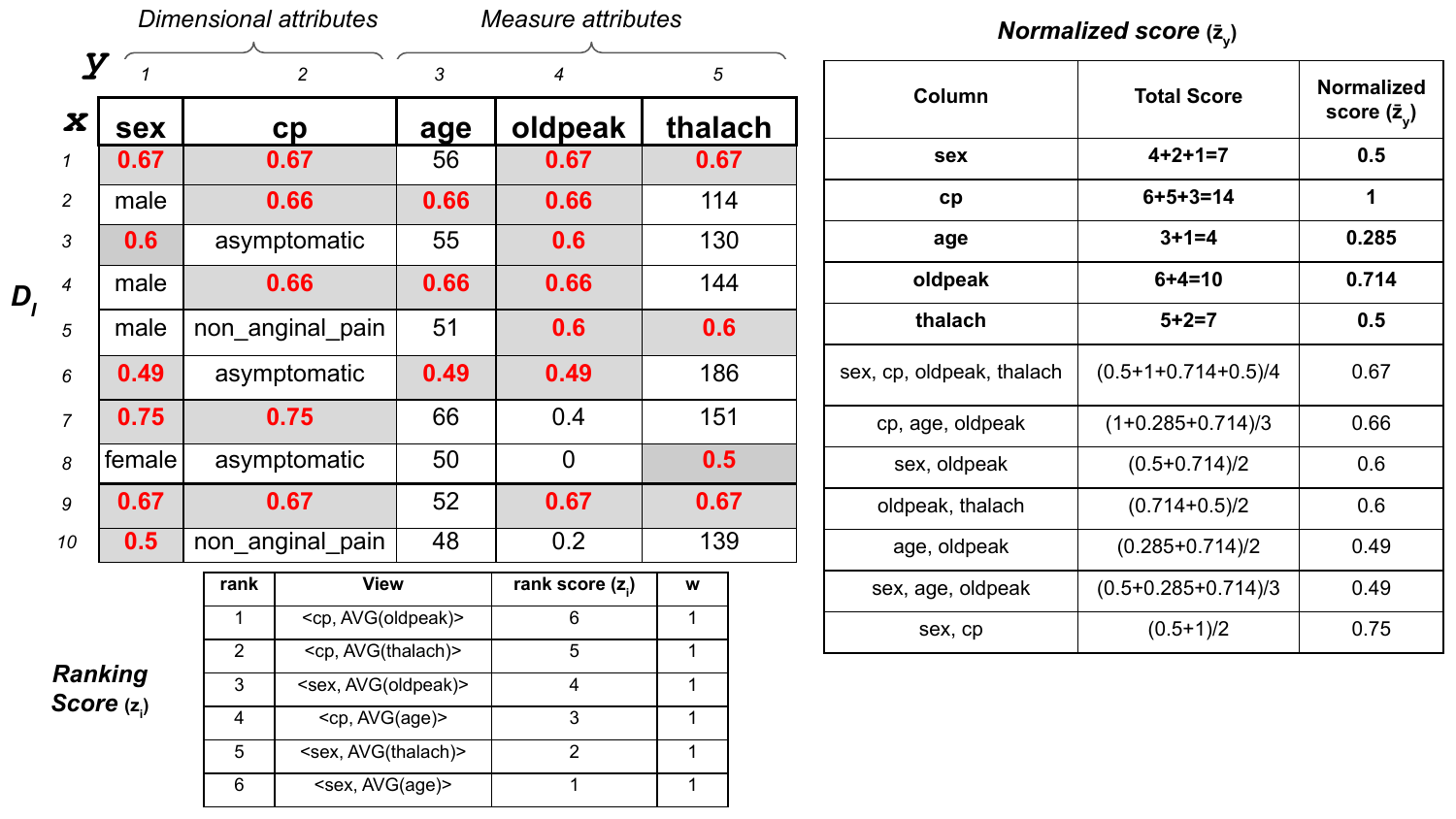}
\caption{VizPut-Ranking score illustration}
\label{fig:pbrank_data}
\end{figure}

Consider a data analyst working with incomplete data $D_I$ and an imputation budget $g$. Initially, a temporary insight, \textit{temp-rank}, is derived directly from $D_I$. Subsequently, each view $V_i$ is assigned a ranking score $z_i$, reflecting its importance score, and \textit{temp-rank} is sorted in descending order based on $z_i$ (refer to the Ranking Score table in Figure~\ref{fig:pbrank_data}). Following this, the ranking scores of the visualizations are mapped to the incomplete data $D_I$.

To map the ranking score of a visualization to the priority score of a missing cell, each visualization is broken down to the column $y$ level. As depicted in the Normalized Score table of Figure~\ref{fig:pbrank_data}, both dimensional and measure attributes occur multiple times in \textit{temp-rank}. Thus, the ranking score of a column $\hat{z}_y$ is computed by summing the rank score $z_i$ and normalizing by the highest column's ranking score (alternative normalization functions, such as max-min normalization, can be employed). Normalization ensures that the priority score is bounded by 1.

Once each dimension (i.e., column) has a normalized score, denoted as $\hat{z}_y$, the priority score for each missing cell $C_{x,y}$ in $D_I$ is calculated based on the value of $\hat{z}_y$ and the number of missing cells in row $x$, represented by $R$. This can be defined as follows:

If there is only a single missing cell in row $x$, then:
\begin{equation}
\hat{z}_y = \frac{\sum_{y=1}^{|y|} z_i}{max(\sum_{y=1}^{|y|} z_i)}, \quad R(C_{x,y}) = \hat{z}_y
\label{eq:ranking_aware_1}
\end{equation}

If multiple missing cells are in row $x$, then:

\begin{equation}
R(C_{x,y}) = \frac{\sum_{y=1}^{|y|} \hat{z}_y}{|y|}
\label{eq:ranking_aware_m}
\end{equation}.

Hence, we introduce \textit{VizPut-Ranking}, a method that directly generates \textit{temp-rank} and uses it to map the priority of the missing cells to be imputed, even though \textit{temp-rank} may be misleading. The priority function $P(C_{x,y})$ for a missing cell in the \textit{VizPut-Ranking} method is defined as:

\begin{equation}
P(C_{x,y}) = R(C_{x,y})
\label{eq:ranking_aware_final}
\end{equation}

In this equation, $P(C_{x,y})$ represents the priority score of $C_{x,y}$, while $R(C_{x,y})$ denotes the normalized ranking score of the visualization related to $C_{x,y}$ (as described in Equations~\ref{eq:ranking_aware_1} and \ref{eq:ranking_aware_m}).

To provide a clear illustration of the proposed approach, consider the example of cell \(C_{10,1}\), located in column 1, row 10, with a priority score of 0.5. This score arises because there is only one missing cell in row 10. Thus, the priority score is calculated by summing \(z_i\) from the "sex" column. Three visualizations include the "sex" attribute, with values \(z_3 = 4\), \(z_5 = 2\), and \(z_6 = 1\), resulting in a total of 7. To normalize the value to 1, this total is divided by the maximum total score \(\hat{z}_y\), which is 14.

As another example, consider cell \(C_{1,y}\) in the first row, which has four missing cells corresponding to the columns "sex", "cp", "oldpeak", and "thalach". The priority scores of these missing cells are first calculated based on the column scores (refer to the normalized score table \(\hat{z}_y\)), where the columns "sex", "cp", "age", "oldpeak", and "thalach" have scores of 7, 14, 4, 10, and 7, respectively. Consequently, to calculate the priority score of \(C_{1,y}\), the scores of the columns "sex", "cp", "oldpeak", and "thalach" are summed and then divided by the number of those columns. This results in a \(\hat{z}_y\) score for \(C_{1,y}\) of $\frac{0.5+1+0.714+0.5}{4} = 0.67$

\eat{
To provide a clear illustration of the proposed approach, c consider the first row $C_{1,y}$ of the incomplete dataset $D_I$. This row exhibits a single complete cell (i.e., \textit{age}) alongside four missing cells (i.e., \textit{sex, cp, oldpeak, thalach}). To determine the priority score $R(C_{x,y})$ for the missing cells within this row, we define $R(C_{x,y})$ as the sum of the normalized scores $\hat{z}$ for the missing cells (\textit{sex, cp, oldpeak, thalach}), divided by the quantity of missing cells present in this row (Equation~\ref{eq:ranking_aware_final}). This computation is illustrated in the Normalized Score table of Figure~\ref{fig:pbrank_data}, specifically in row 6. As a result, the priority score for all missing values in row $1$ is computed as $0.67$. The $R(C_{x,y})$ score corresponding to each missing cell in $D_I$ can be observed in Figure~\ref{fig:pbrank_data}.
}

As depicted in Figure~\ref{fig:pbrank_data}, given a single imputation budget constraint, the missing cells warranting prioritization include $C_{7,1}$ and $C_{7,2}$. Both of these cells exhibit the highest priority score, $R(C_{x,y})$, valued at $0.75$. In instances where multiple missing cells share the highest score, a random selection process is employed. A detailed account of this methodology is provided in Algorithm~\ref{VizPut_Ranking_Aware_algorithm}.

\setlength{\textfloatsep}{0pt}
\begin{algorithm}[t]
\DontPrintSemicolon
\SetKwBlock{Begin}{}{end}
	\KwIn{Incomplete data $D_I$, imputation budget $g$, views $k$, ranking weight $w$ }
	\KwOut{Imputed data $D_f$}  
	 \SetKwFunction{FMain}{VizPutBasedOnTemporaryRanking}
     $\mathbb{V} \leftarrow $ generate views from $D_I$, $ k \leftarrow 10$\;
    $ C_{x,y}  \leftarrow $ a missing cell(row $i$, column $j$) on $ D_I $\;   
    $ w  \leftarrow $\{\textit{1:VizPut-Ranking, 2:VizPut-Ranking(w)}\}$ $\;
    \SetKwProg{Fn}{Function}{:}{}
    \Fn{\FMain{$D_I, k, w, g$}}{

        \If{$ w = 2 $ }
            { 
            $S \leftarrow $ get top-k views from $(\mathbb{V})$\;
            generate ranking score of each view (Figure~\ref{fig:pbrank_data})\;
            map ranking score ($S$ $\times$ linear increase, $\mathbb{V}\backslash S$ $\times$ linear decrease)\;
            calculate $\hat{R}(C_{x,y})$, normalize and distribute to missing $C_{x,y}$}
        \Else{
            map ranking score $\mathbb{V} \times 1$, calculate $\hat{R}(C_{x,y})$ normalize and distribute to missing $C_{x,y}$
        }
    	$C_{S} \leftarrow $ get all highest score $C_{x,y}$ with the size of $g$\;
        \textbf{return} $ C_{S} $ 
        
    }
    \textbf{End Function}
	
		$ C_{S} \leftarrow VizPutBasedOnTemporaryRanking(D_I, k, w, g))$\;
		$ D_f.impute\left(C_{S}\right)$\;

	return $D_f$
	\caption{VizPut-Ranking Algorithm}
	\label{VizPut_Ranking_Aware_algorithm}
\end{algorithm}

In the \textit{VizPut-Ranking} approach, an advantage emerges through the strategic allocation of imputation resources towards missing cells associated with the prospective top-$k$ insights. Nevertheless, this approach is not devoid of drawbacks. Owing to the fact that the \textit{temp-rank} and candidate top-$k$ insights are derived from the incomplete dataset $D_I$, the \textit{temp-rank} is susceptible to inaccuracies, such as incorrect ranking order, as a result of the presence of missing values within $D_I$.

This phenomenon consequently results in an increased probability of imputing missing cells associated with the top-$k$ on the \textit{temp-rank}. As the \textit{VizPut-Ranking} approach calculates the priority score of a missing cell based on the ranking score of its corresponding visualization, it may lead to the imputation of missing cells linked to top visualizations that consistently maintain their high-ranking positions, thereby consuming the imputation budget $g$ without significantly enhancing the recommendation accuracy.
Additionally, there exists a possibility that certain visualizations, initially positioned below the top-$k$, may exhibit a high potential for promotion to the top-$k$ set, while some visualizations nearing the bottom of the top-$k$ set could face exclusion. This observation further underscores the limitations of the \textit{VizPut-Ranking} technique in the context of achieving optimal imputation priority outcomes.

In order to address this challenge, we propose an alternative weighted schema for visualizations within the \textit{temp-rank}, denoted as \textit{VizPut-Ranking(w)}. This model accommodates varying weights based on the position or ranking of the visualization. The details of this approach will be discussed in the subsequent section.

\begin{figure}[t!]
\centering
\includegraphics[width=5in]{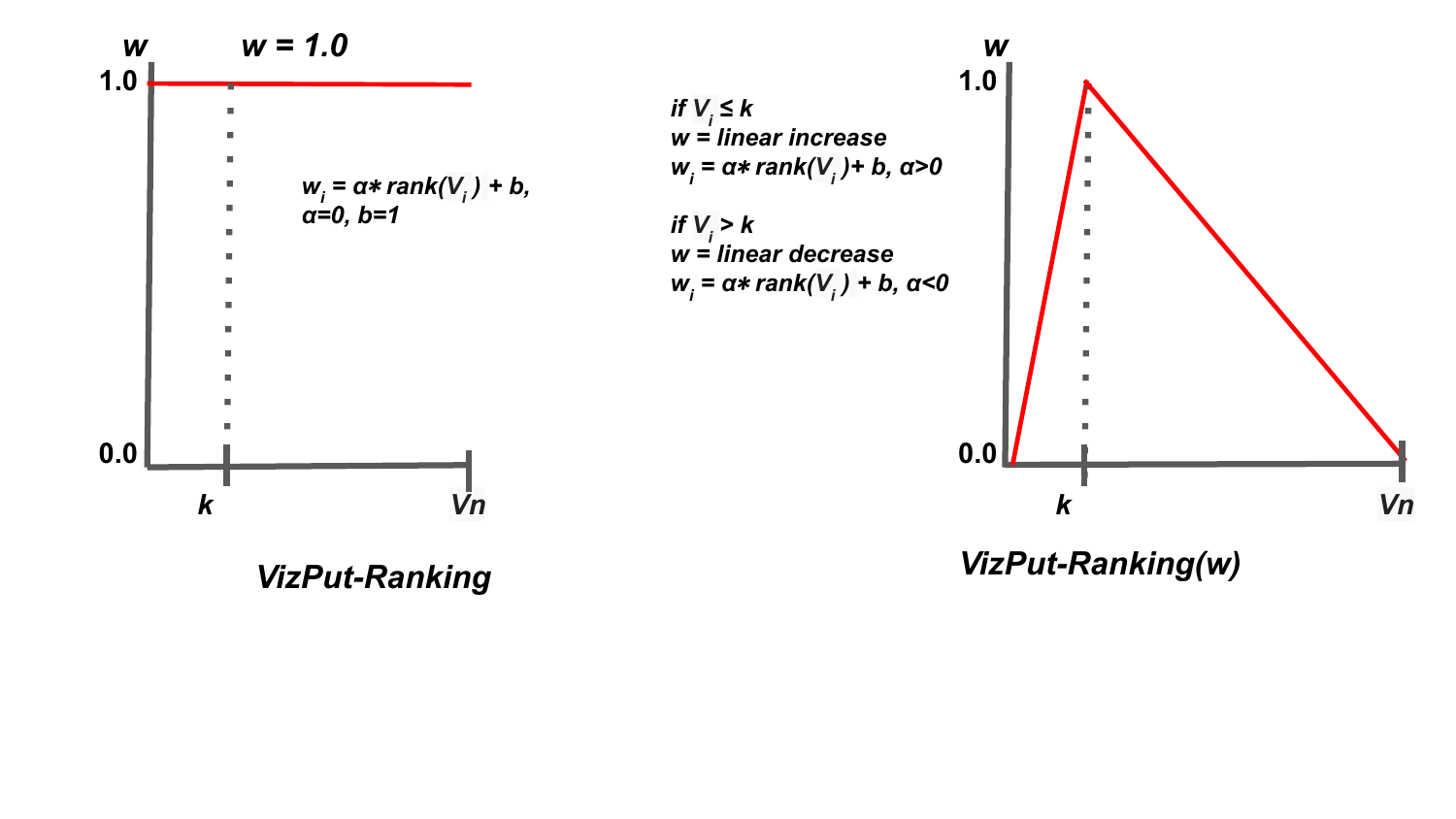}
\vspace{-60pt}
\caption{VizPut-Ranking vs VizPut-Ranking(w) illustration}
\label{fig:VizPut-Ranking_weight}
\end{figure}

\subsubsection{Weighted VizPut-Ranking}
\label{subsection:VizPut-Ranking_wb}

Consider the scenario in which a data analyst is dealing with incomplete data \(D_I\) and has an imputation budget of \(g\). Initially, a provisional insight, termed \textit{temp-rank}, is extracted directly from \(D_I\). Subsequent to this, each view \(V_i\) is attributed a ranking score \(z_i\), representing its significance. The \textit{temp-rank} is then organized in descending order according to the \(z_i\) values (please refer to the Ranking Score table in Figure~\ref{fig:pbrank_data}). In the context of the \textit{VizPut-Ranking} method, the weighting of the visualization is consistently maintained at a value, indicating that no differential weighting is applied; specifically, \(w_i=1.0\) where \(w_i\) is the weight of each $V_i$ and \(z_i = z_i \times w_i\). As a result, the imputation priority is exclusively determined by the ranking score of the \textit{temp-rank}, generated based on the incomplete dataset \(D_I\). However, the reliability of this ranking could be subject to scrutiny as it is derived from the incomplete dataset \(D_I\).

In order to enhance the \textit{VizPut-Ranking} method, we introduce an extension that assigns weight to each visualization within the \textit{temp-rank}, designated as \textit{VizPut-Ranking(w)}. As a result, the imputation priority is determined by the product of the ranking score of visualization $z_i$ and weight $w_i$. In the \textit{VizPut-Ranking(w)} model, it is posited that top visualizations, such as top-1 or top-2, typically remain at the top, and therefore, should be assigned a lower weight. Conversely, visualizations situated near the end of the top-$k$ set ought to be allocated a higher weight due to their potential for rank fluctuation.

In this model, the weight $w_i$ follows a linearly increasing trend if $V_i \leq k$ and a linearly decreasing trend if $V_i>k$. Both linear increase and decrease conform to the linear model function. The linear increase in weight is defined as:

\begin{equation}
w_i = \alpha \times \text{{rank}}(V_i) + b
\end{equation}

where $\alpha > 0$. At the same time, the decrease in weight is defined as:

\begin{equation}
w_i = \alpha \times \text{{rank}}(V_i) + b
\end{equation}

where $\alpha < 0$. The overall graph displays an inverted-U shape, as depicted in Figure~\ref{fig:VizPut-Ranking_weight}. The figure illustrates the differing weight distributions: \textit{VizPut-Ranking} employs a constant weight, while \textit{VizPut-Ranking(w)} manifests an inverted-U weight distribution.

The mapping method from the ranking results to cells in \textit{VizPut-Ranking(w)} is done using the same method as in \textit{VizPut-Ranking}. However, to calculate $\hat{z}_y$, the $z_i$ score is multiplied by $w_i$ as defined in:

\begin{equation}
\hat{z}_y = \frac{\sum_{y=1}^{|y|} z_i \times w_i}{\max(\sum_{y=1}^{|y|} z_i \times w_i)}
\label{eq:ranking_aware_inverted_u}
\end{equation}

In the \textit{VizPut-Ranking(w)} model, the coefficients \(\alpha\) and \(b\) are utilized to ensure that the weights adhere to a linear model. Since the values of \(\hat{z}_y\) are normalized, the choice of \(\alpha\) and \(b\) is quite flexible, acting as scaling factors. We can use any values of \(\alpha\) and \(b\) that meet the criteria: \(\alpha > 0\) for a linear increase and \(\alpha < 0\) for a linear decrease. The model will yield equivalent results post-normalization, as these coefficients primarily serve to shape the initial distribution of the weights. Consider the scenario where the number of visualizations, denoted as $\mathbb{V}$, equals 100 and let $k = 10$. For the phase of linear increase, we assign $\alpha = k$ and $\beta = 0$. Consequently, the values of $w_i$ range from 10 to 100, increasing linearly when $k$ varies from 1 to 10. Subsequently, in the phase of linear decrease, we allocate values $\alpha = -1$ and $\beta = |\mathbb{V}| + k$, resulting in $\beta = 110$. By establishing these parameters, the values of $w_i$ decrement linearly, commencing at 99 when $k=11$, and terminating at 10 when $k=100$. This results in a U-shaped distribution of the values of $w_i$.

Finally, the priority function for the \textit{VizPut-Ranking(w)} approach is defined as:

\begin{equation}
P(C_{x,y}) = \hat{R}(C_{x,y})
\end{equation}

where $\hat{R}(C_{x,y})$ represents the weighted ranking score of the missing cell.

\eat{
In \textit{VizPut-Ranking wb constant-decrease}, the imputation priority is based on both the ranking score of the visualization $\bar{z}$ and the weight $w$. In this model, the visualizations in the top-$k$ set are assumed to have equal weights and are assigned the heaviest weights. The weights of visualizations after the top-$k$ decrease linearly following a linear-decrease function. The weight $w$ is defined as follows: 1) $w=1.0$ if $V_i\leq k$; 2) a decreasing weight if $V_i>k$. The constant weight follows the linear model $w=\alpha\times rank(V_i)+b$, where $\alpha=0$ and $b=1$, while the decreasing weight follows $w=\alpha\times rank(V_i)+b$, where $\alpha<0$. For example, let $k=10$. If $\alpha=-5$ and $b=100$, then $w=-5\times 11 + 100 = 45$ for $V_{11}$, and $w=40, 35, 30, 25$ for $V_{12}, V_{13}, V_{14}, V_{15}$, respectively. $\alpha$ and $b$ can be defined based on user preferences. The contribution score is then normalized. 
}

\subsection{Hybrid approach}
\label{subsection:hybrid}

In the present study, we propose a hybrid approach that integrates the \textit{VizPut-Ranking(w)} and \textit{VizPut-Cell(f)} algorithms with the objective of enhancing the performance in handling incomplete data within visualization recommendation tasks. Initially, this hybrid strategy generates a temporary rank from the incomplete dataset, followed by the computation of the priority score for each missing cell, based on a combination of the \textit{VizPut-Ranking(w)} and \textit{VizPut-Cell(f)} algorithms.

The priority score within the \textit{Hybrid} approach is defined as follows:

\begin{equation}
P(C_{x,y}) = \hat{R}(C_{x,y}) \times N(C_{x,y}) \times F(C_{x,y})
\end{equation}

Here, $N(C_{x,y})$ represents the contribution score of the cell to the recommendation results, $F(C_{x,y})$ is the fairness score, and $\hat{R}(C_{x,y})$ denotes the weighted ranking score of the visualization associated with the missing cell. The effectiveness of our hybrid approach is evaluated and presented in Section~\ref{sec:experiment_evaluation_vizput}.

\subsection{VizPut Optimization}
\label{subsection:efficiency_comparison}

\begin{figure}[t!]
\centering
\includegraphics[width=5in]{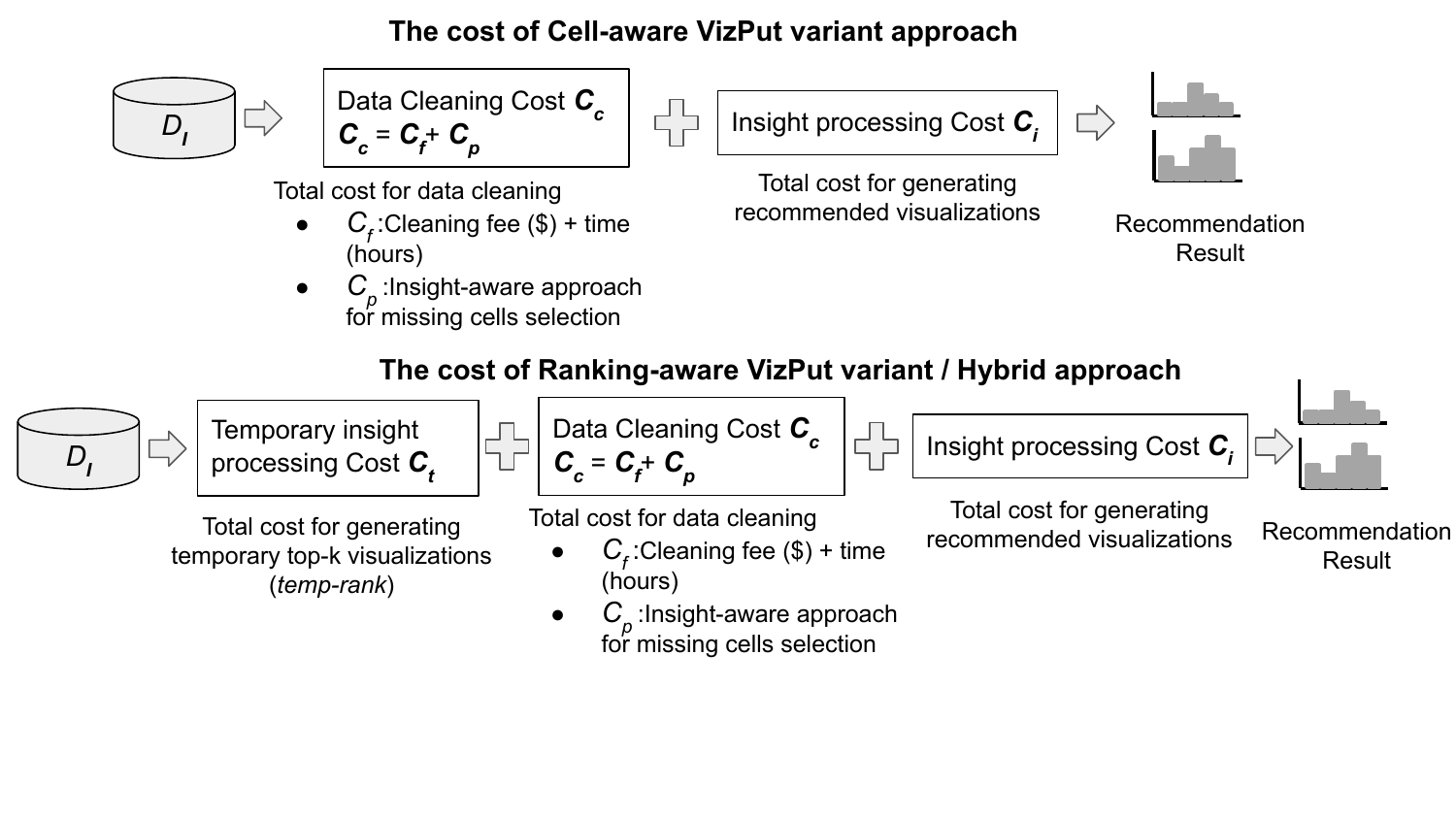}
\vspace{-40pt}
\caption{The cost comparison of Cell-aware VizPut, Ranking-aware VizPut, and Hybrid}
\label{fig:efficiency_comparison}
\end{figure}

To explain the cost of VizPut, consider Figure~\ref{fig:efficiency_comparison}, which compares the overall cost of \textit{Cell-aware VizPut}, \textit{Ranking-aware VizPut}, and \textit{Hybrid}. As shown in Figure~\ref{fig:efficiency_comparison}, the total cost to generate the recommendation results is the sum of the data cleaning cost $C_c$ and the insight processing cost $C_i$.  Specifically, the data cleaning cost $C_c$ comprises two components:

\begin{enumerate}
    \item Cleaning fees $C_f$: These might entail both monetary and temporal expenditures, especially if the data analyst employs experts to rectify incomplete data.
    \item Computation of the priority score for the missing cell $C_p$: This represents the selection of the missing cell. The cost varies depending on whether we utilize \textit{Cell-aware VizPut}, \textit{Ranking-aware VizPut}, or \textit{Hybrid}.
\end{enumerate}

The discrepancy in $C_p$ across the different VizPut approaches arises from the distinct priority function calculations inherent to each approach. For instance, the calculation of the priority score for missing cells in \textit{Hybrid} is given by $P(C_{x,y}) = \hat{R}(C_{x,y}) \times N(C_{x,y}) \times F(C_{x,y})$, whereas in \textit{VizPut-Ranking} it is $P(C_{x,y}) = R(C_{x,y})$. Consequently, the cost of missing cell selection $C_p$ for Hybrid will be greater than that for VizPut-Ranking, as Hybrid requires calculations considering both fairness and the impact of the cell.

In addition, the insight processing cost $ C_i $ encompasses all processes required to generate recommended visualizations. According to Figure~\ref{fig:efficiency_comparison}, the main difference in cost between our proposed approaches lies in the number of times insights are generated. For instance, the \textit{Cell-aware VizPut} variants generate insights only once, as they follow the \textit{impute-first-insight-next} approach, where recommended visualizations are produced after the data cleaning process is completed. In contrast, the \textit{Ranking-aware VizPut} variants generate insights twice, as they follow the \textit{insight-first-impute-next} approach. In this approach, the top-k candidate insights, referred to as \textit{temp-rank}, are generated first, incurring an additional cost $C_t$, followed by imputation based on the \textit{temp-rank}, and finally, recommended visualizations are produced $C_i$.  Since the hybrid approach constructs the \textit{temp-rank} from incomplete data prior to any data cleaning, analogous to \textit{Ranking-aware VizPut}, its total cost for generating recommendations is akin to that of \textit{Ranking-aware VizPut}. However, the data cleaning cost \( C_c \) for the \textit{Hybrid} approach diverges from \textit{Ranking-aware VizPut} due to variations in \( C_p \).

To avoid fully computing \( C_i \), especially for the \textit{Ranking-aware VizPut} variants and \textit{Hybrid} — given that both approaches first generate \textit{temp-rank} which incurs a cost \( C_t \) comparable to \( C_i \) — we propose several optimization strategies to selectively determine which visualizations should be regenerated a second time:

\begin{enumerate}
 \item \textbf{no-opt / baseline}: This approach involves regenerating all visualizations a second time, doubling the expense of insight generation. While this baseline method maximizes effectiveness by regenerating all visualizations post-imputation, it comes at a substantial cost.

    \item \textbf{top-k}: Only visualizations of size \( k \), as defined by the user, are regenerated a second time. Thus, the top-k from \textit{temp-rank} will be regenerated. The order of visualizations might change following by the improvement of effectiveness, but those visualizations beyond the top-k will not be re-executed, even though they might have a chance to join the top-k following imputation.
    \item \textbf{top-k highest imputed}: Only the top \( k \) visualizations that received the highest number of imputed cells will be regenerated a second time.
    \item \textbf{top-k + top-k highest imputed}: This method combines the previous two. The top-k visualizations often overlap with those that receive the highest imputation. For generating the final recommendation results, the visualizations from both categories are combined, and the top \( k \) from this combined set are selected and recommended to the user. In this work, we rely on this method for both \textit{Ranking-aware VizPut} variants and \textit{Hybrid}.
\end{enumerate}

The experimental results related to these optimizations can be found in Section~\ref{sec:experiment_evaluation_vizput}.

\vspace{-10pt}
\section{Experimental Evaluation}
\vspace{-10pt}
\label{sec:experiment_evaluation_vizput}

In this section, we first discuss our experimental testbed (Sec.~\ref{subsection:testbed}). Then, we present and discuss our experimental evaluation. (Sec.~\ref{subsection:experiment_result_vizput}). 

\subsection{Experimental Testbed}
\label{subsection:testbed}

\noindent{\textbf{Datasets: }}We conduct experiments on the following datasets: (1) The Cleveland Heart Disease dataset, which comprises 8 dimensional attributes, 6 measures \cite{HeartDisease}. (2) The New York Airbnb dataset, which comprises 4 dimensional attributes, 4 measures, and 30,249 tuples \cite{airbnb}. (3) The Diabetes 130 US Hospital dataset, which consists of 14 dimensional attributes, 13 measures, and 100,000 tuples \cite{DiabDataset}. Although we perform experiments on all three datasets, the Cleveland Heart Disease dataset is used as the default dataset for presenting results in this paper due to space limitations. The overall priority function $P(C_{x,y})$ can be expressed as a single function or as a combination of two or more functions, utilizing multiplication as the combining operator. Alternatively, we can resort to using addition, which would yield comparable results due to the fact that the priority scores of missing cells are eventually ranked. The selection of these cells is then based on descending order of these priority scores. 

\eat{
Nevertheless, in scenarios where the dimensionality of the dataset is significantly high, relying on addition might pose challenges. This is because some normalization procedures are dependent on the number of dimensions, which could potentially skew the prioritization process. To standardize the values of each priority function, we bound their values to the range of 0 and 1, and normalize them using standard normalization functions such as max normalization $\bar{z} = \frac{z}{\max(z)}$ or max-min normalization $\bar{z} = \frac{z - \min(z)}{\max(z) - \min(z)}$.}

\noindent{\textbf{Incomplete data: }}
We simulate missing data using the completely at random (MCAR) method with respect to the entire dataset. In this experiment, we create an incomplete version of the data, $D_I$, from the original dataset $D$. In order to avoid bias, 100 versions of $D_I$ with different random missing seeds are generated. 

\noindent{\textbf{Imputed data: }}
In our research, we start with a clean dataset and introduce missing values, while retaining the ground truth data. This approach allows us to evaluate the effectiveness of our priority function in determining which missing cells should be imputed first. The imputed values are considered 100\% correct, as they are based on the ground truth data. 

\noindent{\textbf{Effectiveness Metrics: }}
We employ Jaccard and Rank Biased Overlap (RBO) metrics from our previous work \cite{QualityMatters} to evaluate the effectiveness of our proposed approaches. These metrics are utilized to compare the recommended visualizations generated from incomplete data with those generated from imputed data using our proposed methods.

\noindent{\textbf{Default parameters: }}The default parameters used in our evaluation are $k = 10$, with $20\%$ of missing data, a maximum imputation budget of 10\% relative to the number of missing cells, and effectiveness measurements of Jaccard and RBO. The default data cleaning method is \textit{ignore cell}, and the default dataset is the Cleveland Heart Disease dataset. The final results are the average of 100 versions of $D_I$ and are presented with a confidence interval of $CI = 0.95$. Common aggregate functions such as {\ttfamily AVG, SUM, MAX, MIN} are used. We utilized various query predicates \( T \) in the experiment to ensure the reliability of the recommendation results.

\eat{
The query predicates $T$ used in the experiment include the following: for the Heart Disease dataset, the default query predicate is $D_Q:$ \textit{disease = Yes} and $D_R:$ \textit{whole data}; for the Airbnb dataset, $D_Q:$ \textit{room type = Entire home/apt} and $D_R:$ \textit{whole data}; and for the Diabetes dataset, $D_Q:$ \textit{readmitted = No} and $D_R:$ \textit{whole data}. 
}

\noindent{\textbf{The list of implemented algorithms and their associated priority functions: }}

\begin{itemize}[noitemsep,topsep=0pt]

\item \textit{No Imputation, Random selection imputation, Fairness imputation}: Baseline refers to Sec.~\ref{subsubsection:baseline_vizput}

\item \textit{Cell-aware VizPut}
    \begin{itemize}
    \item \textit{VizPut-Cell}: \( P(C_{x,y}) = N(C_{x,y}) \)
    \item \textit{VizPut-Cell(f)}: \( P(C_{x,y}) = N(C_{x,y}) \times F(C_{x,y}) \)
    \item \textit{VizPut-Cell(v)}: \( P(C_{x,y}) = N(C_{x,y}) \times V(C_{x,y}) \)
    \item \textit{VizPut-Cell(f, v)}: \( P(C_{x,y}) = N(C_{x,y}) \times F(C_{x,y}) \times V(C_{x,y}) \)
    \end{itemize}

\item \textit{Ranking-aware VizPut}
    \begin{itemize}
    \item \textit{VizPut-Ranking}: \( P(C_{x,y}) = R(C_{x,y}) \)
    \item \textit{VizPut-Ranking(w)}: \( P(C_{x,y}) = \hat{R}(C_{x,y}) \)
    \end{itemize}

\item \textit{Hybrid}: \( P(C_{x,y}) = \hat{R}(C_{x,y}) \times N(C_{x,y}) \times F(C_{x,y}) \)

Where $P$ is the overall priority score, $N$ is the contribution score of the cell to the recommendation results, $F$ is the fairness score, $V$ is the number of used views, $R$ is the ranking score of the visualization associated to the missing cell, and $\hat{R}$ is the ranking of the visualization and the weighted ranking associated with the missing cell. In addition, the final recommendation for both \textit{VizPut-Ranking(w)} and \textit{Hybrid} is generated using the top-k + k-highest imputed approach for optimization.

\end{itemize}

\subsection{Experimental Evaluation}
\label{subsection:experiment_result_vizput}

\begin{figure*}[t!]
\centering
	\begin{subfigure}{0.32\textwidth}
		\includegraphics[width=\linewidth]{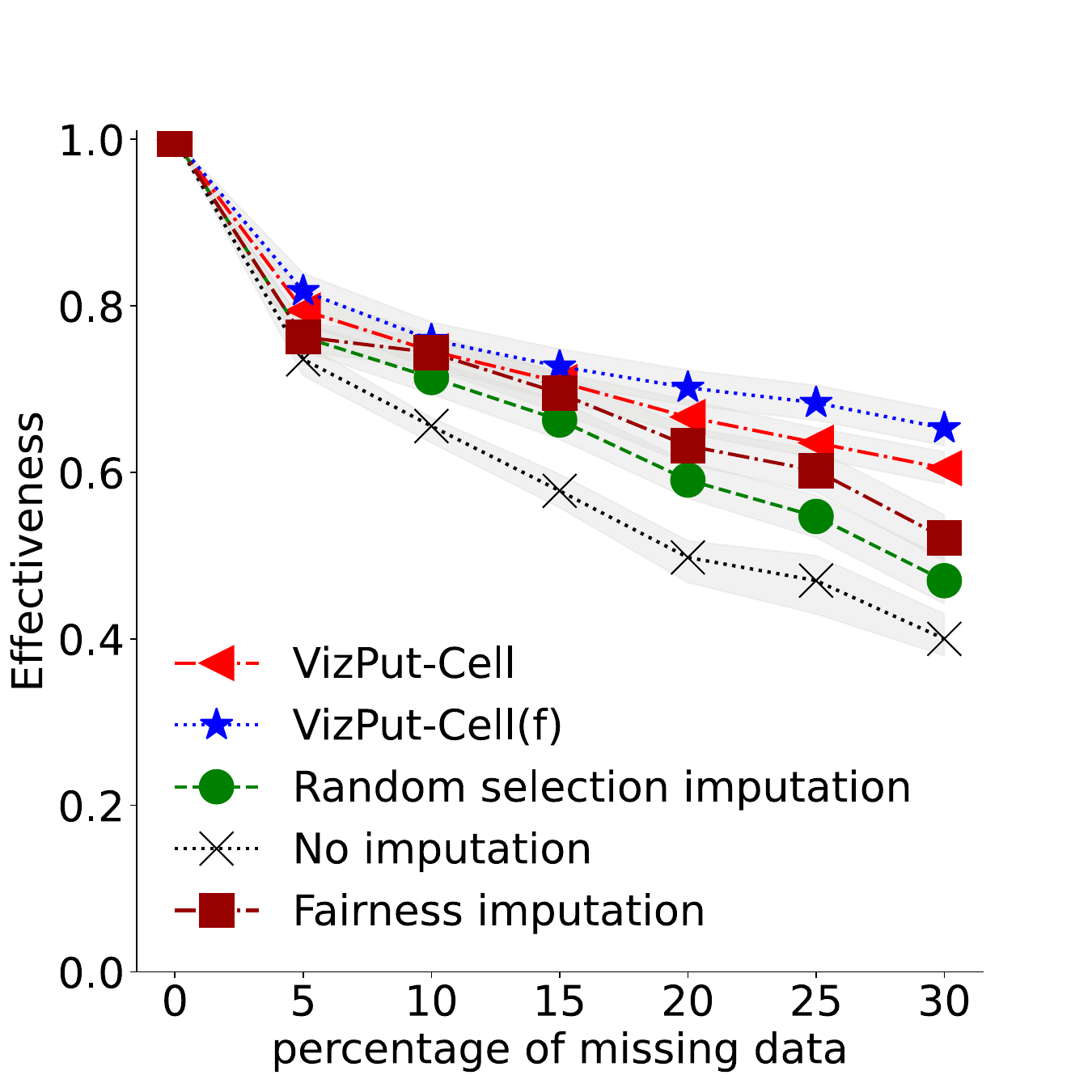}
		\caption{Jaccard} \label{fig:VizPut-Cell_jaccard}
	\end{subfigure}
	\begin{subfigure}{0.32\textwidth}
		\includegraphics[width=\linewidth]{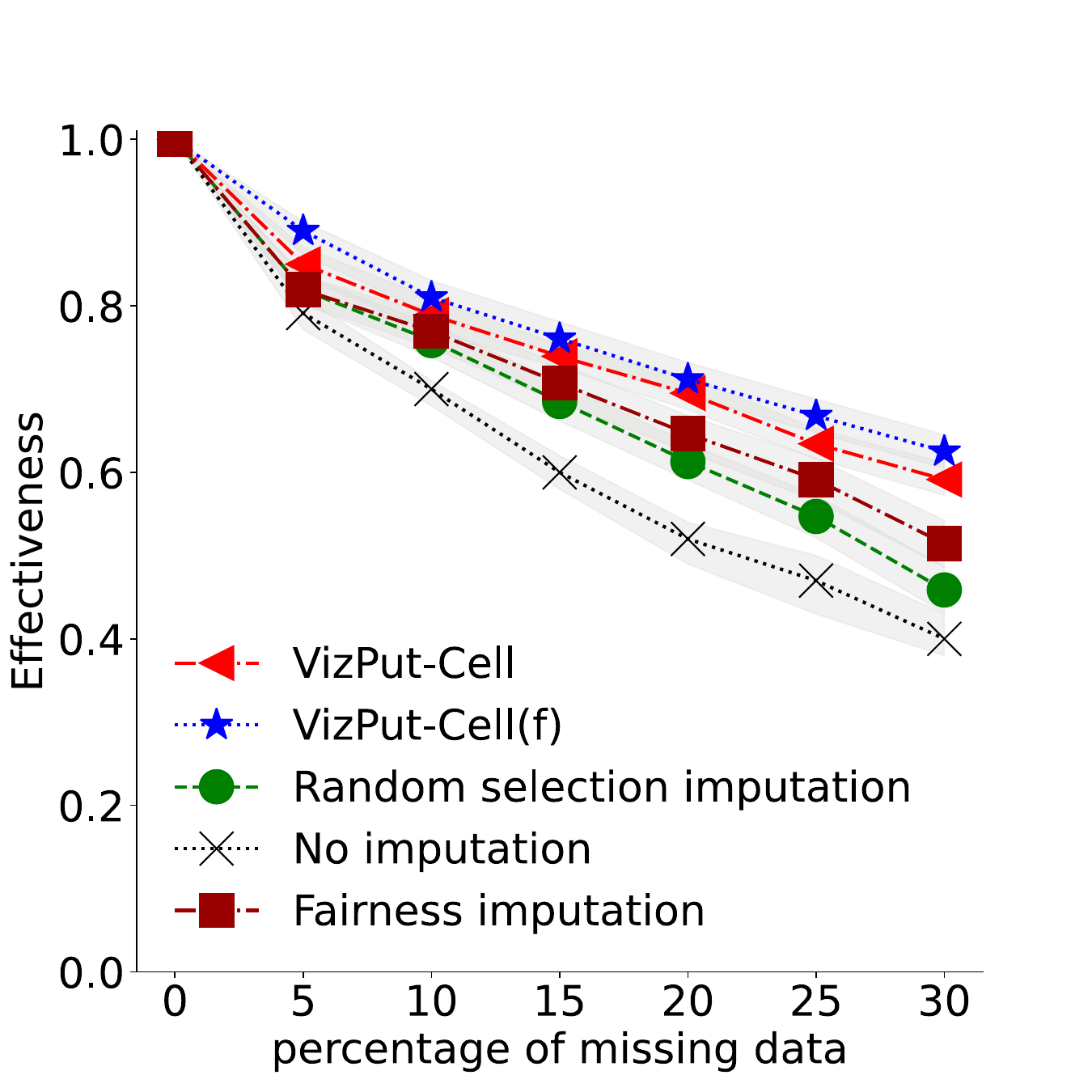}
		\caption{RBO}  

		\label{fig:VizPut-Cell_rbo}
	\end{subfigure}
	\begin{subfigure}{0.32\textwidth}
		\includegraphics[width=\linewidth]{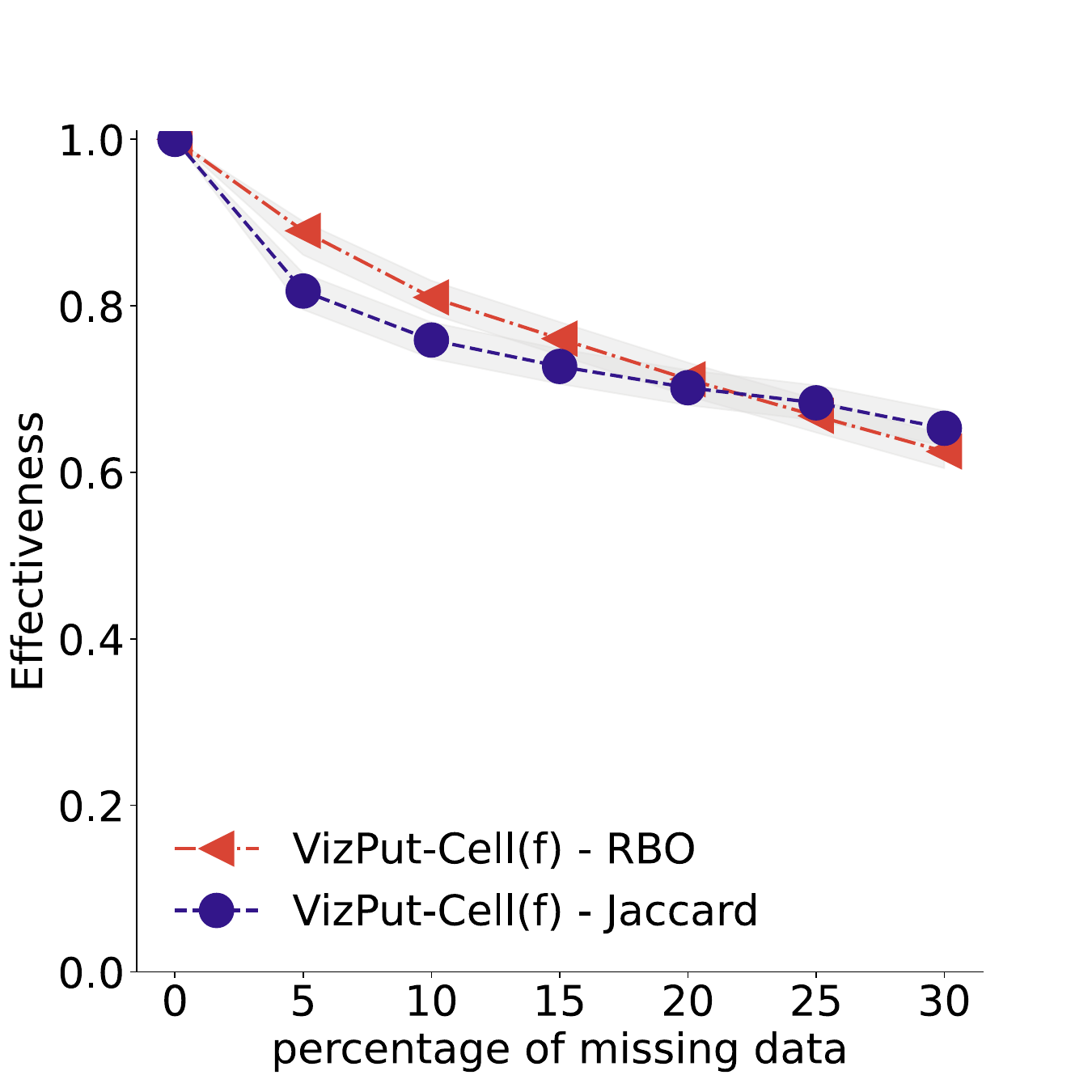}
		\caption{Jaccard vs. RBO}  

		\label{fig:VizPut-Cell_jaccard_vs_rbo}
	\end{subfigure}
	\caption{The performance of \textit{VizPut-Cell} and \textit{VizPut-Cell(f)} compared to baselines on heart disease dataset with different effectiveness measurements}
	\label{fig:VizPut-Cell_performance}
\end{figure*}

\noindent{\textbf{Effectiveness Comparison of VizPut-Cell and VizPut-Cell(f) to Baselines}}. In this experiment, we evaluate the effectiveness of our proposed methods, VizPut-Cell and VizPut-Cell(f), in comparison to the baselines of No Imputation, Random Selection Imputation, and Fairness Imputation. The comparison is illustrated in Figure \ref{fig:VizPut-Cell_performance} using various effectiveness measurements. Both VizPut-Cell and VizPut-Cell(f) surpass the baselines in performance, irrespective of whether the Jaccard or RBO metric is employed for effectiveness evaluation, with VizPut-Cell(f) exhibiting superior performance in relation to VizPut-Cell. The results are derived from experiments conducted on the heart disease dataset, wherein the number of dimensional attributes exceeds the number of measure attributes, resulting in VizPut-Cell(f) outperforming VizPut-Cell (as detailed in Sec.~\ref{sec:insight_aware_approach}).

\begin{figure*}[t!]
\centering
	\begin{subfigure}{0.32\textwidth}
		\includegraphics[width=\linewidth]{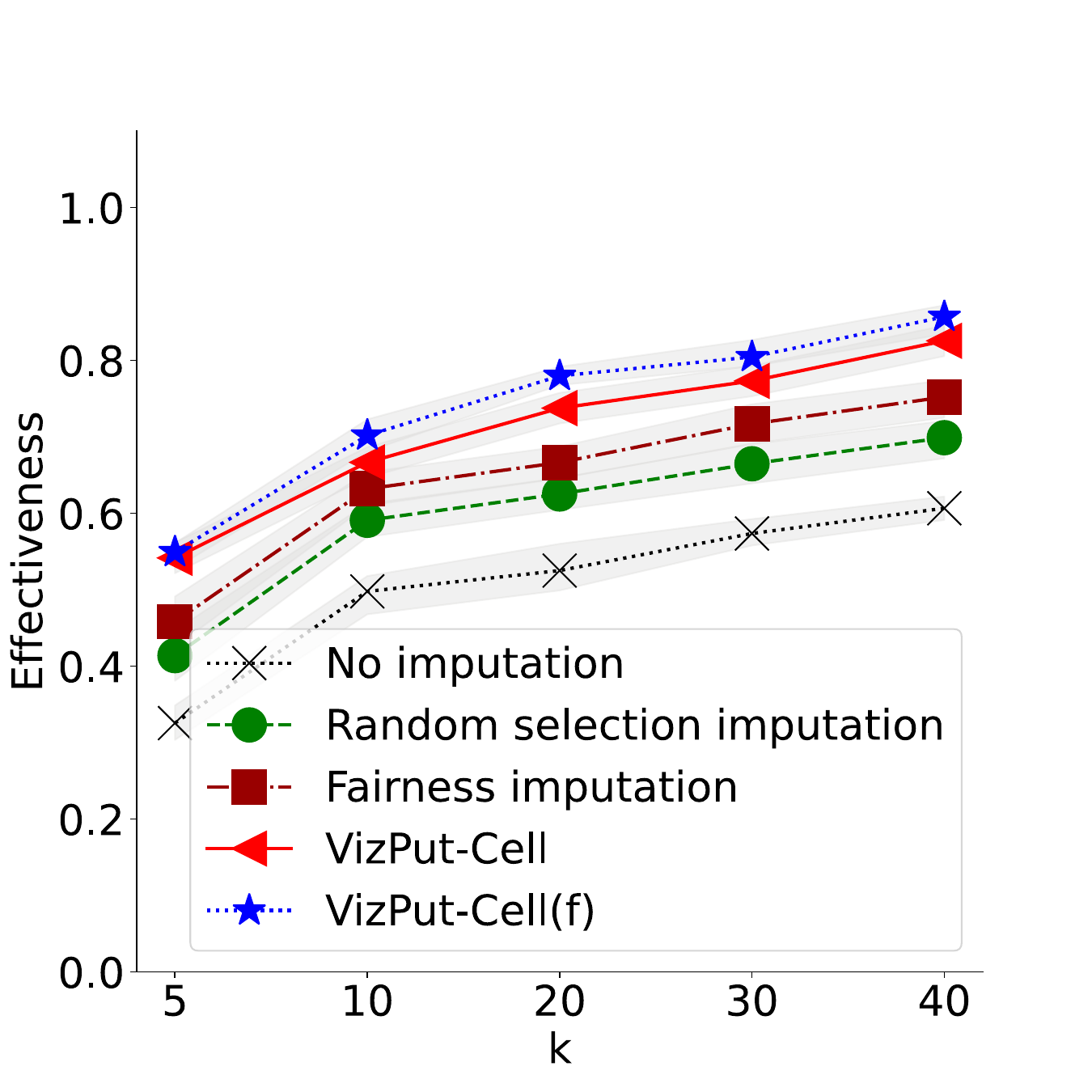}
		\caption{Jaccard} 
		\label{fig:impact_k_jaccard}
	\end{subfigure}
	\begin{subfigure}{0.32\textwidth}
		\includegraphics[width=\linewidth]{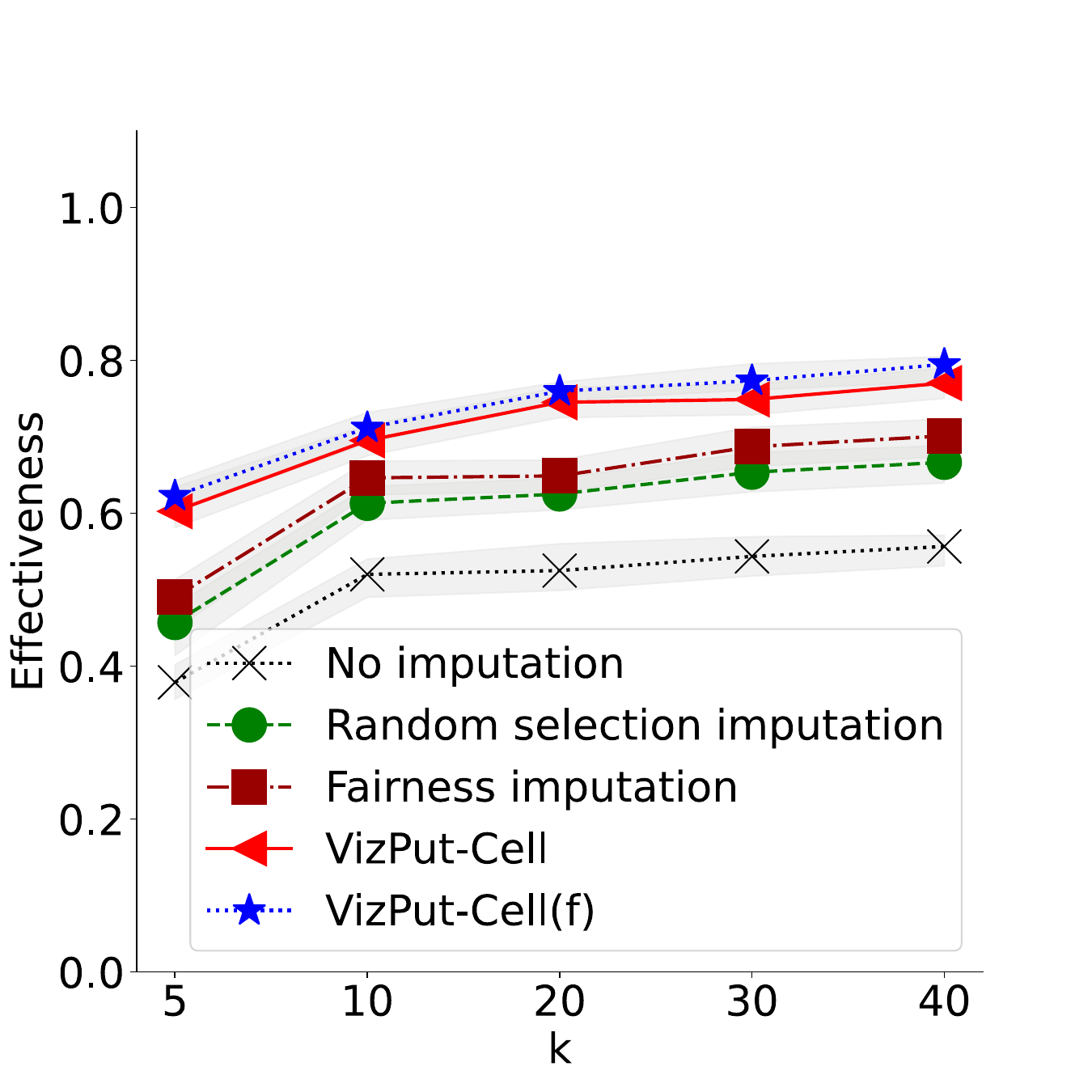}
		\caption{RBO}  
		\label{fig:impact_k_rbo}
	\end{subfigure}
	\begin{subfigure}{0.32\textwidth}
		\includegraphics[width=\linewidth]{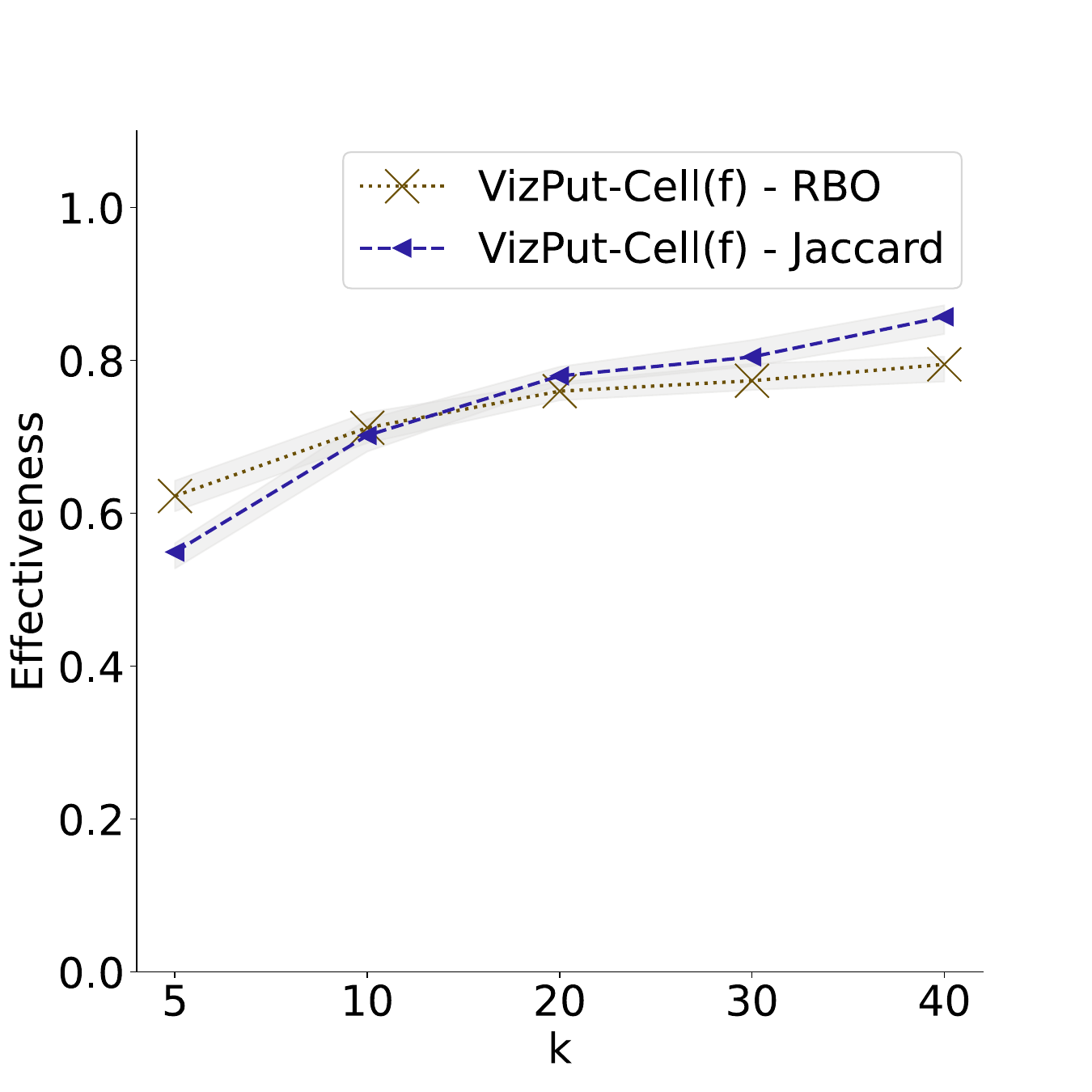}
		\caption{Jaccard vs RBO}  
		\label{fig:impact_k_jaccard_rbo}
	\end{subfigure}
	\caption{Impact $k$ on \textit{VizPut-Cell} over heart disease dataset with different effectiveness measurements} 
	\label{fig:impact_k_VizPut-Cell}
\end{figure*}

\noindent{\textbf{Impact of $k$ on VizPut-Cell and VizPut-Cell(f)}}. The impact of the parameter $k$ on VizPut-Cell and VizPut-Cell(f) is demonstrated in Figure \ref{fig:impact_k_VizPut-Cell}. In general, the effectiveness rises with increasing values of $k$. Figure \ref{fig:impact_k_jaccard_rbo} displays a comparison of the effectiveness using Jaccard and RBO metrics. A crossover between RBO and Jaccard is evident. For small values of $k$ (e.g., $5, 10$), Jaccard underperforms in comparison to RBO. However, for larger $k$ values (e.g., $>20$), Jaccard outperforms RBO. This occurs because higher $k$ values lead to increased effectiveness according to Jaccard, but not necessarily according to RBO. The Jaccard score equals 1 when $k = |\mathbb{V}|$, meaning that the number of $k$ is equal to the number of candidate visualizations. In contrast, the RBO score can only equal 1 if the visualizations within the top-$k$ sets appear in the same order, which is challenging to achieve. Consequently, augmenting the value of $k$ does not guarantee enhanced effectiveness in terms of RBO.

\begin{figure*}[t!]
\centering
	\begin{subfigure}{0.32\textwidth}
		\includegraphics[width=\linewidth]{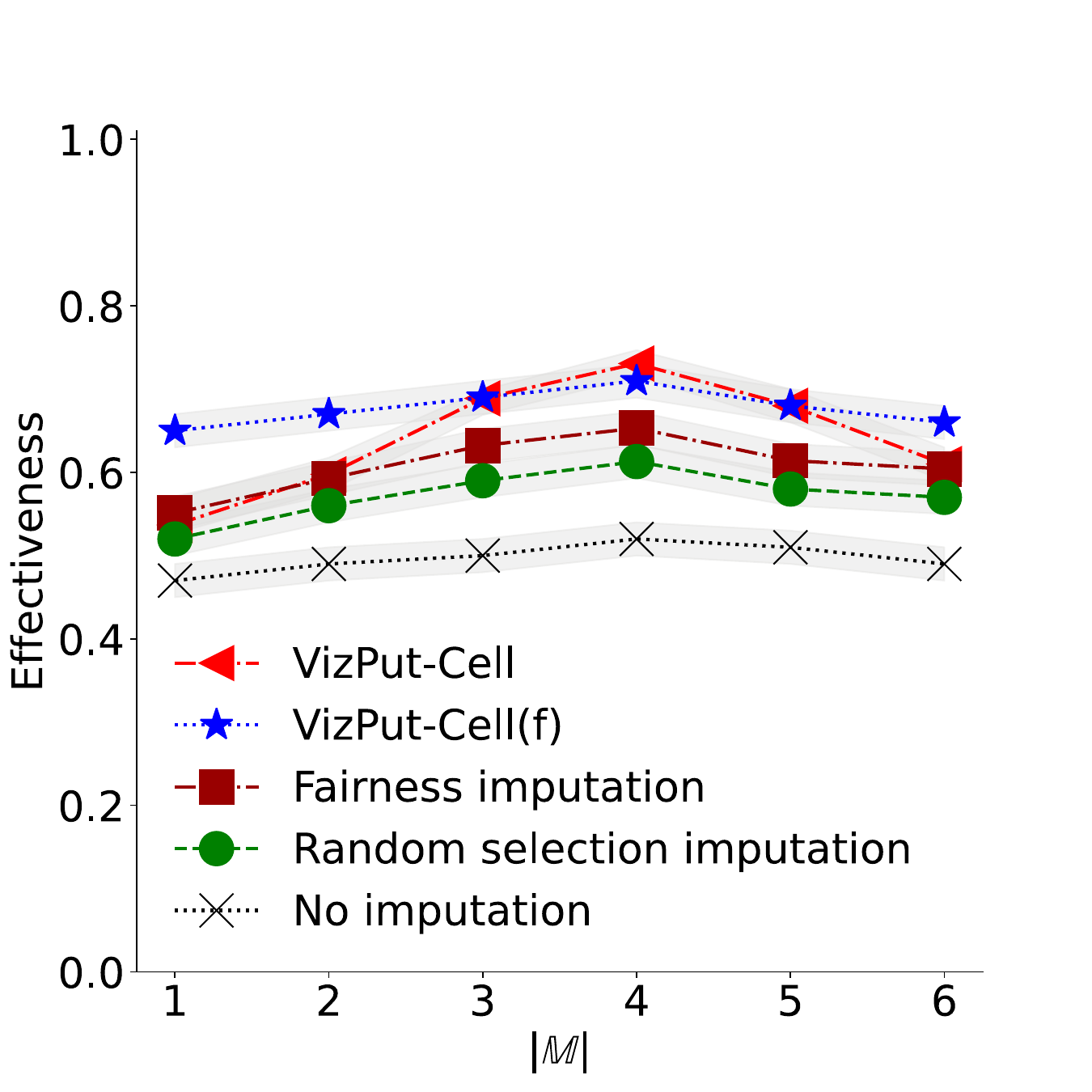}
		\caption{Heart disease dataset, $|\mathbb{A}| = 4$, different $|\mathbb{M}|$} 
		\label{fig:VizPut-Cell_a_m}
	\end{subfigure}
	\begin{subfigure}{0.32\textwidth}
		\includegraphics[width=\linewidth]{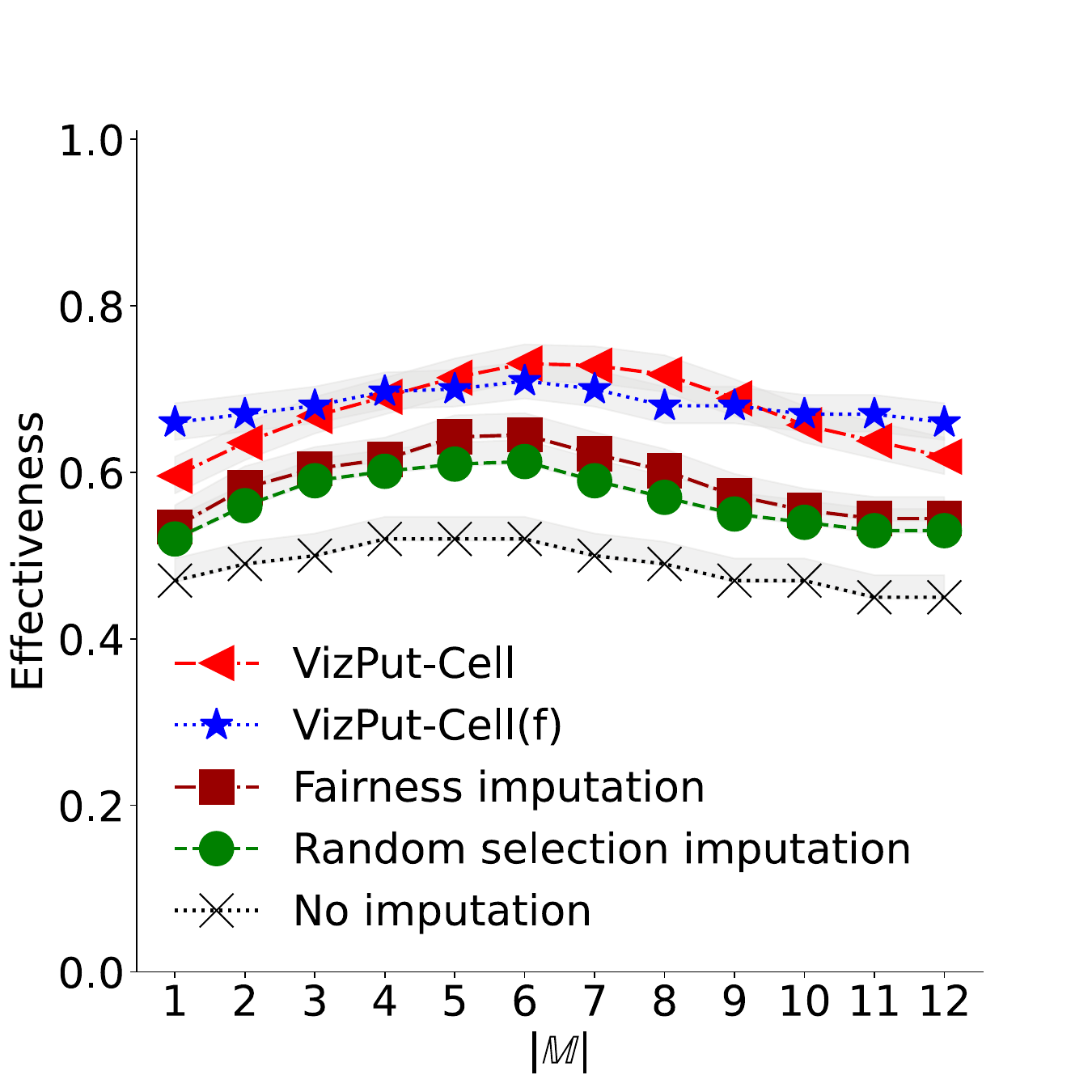}
		\caption{Diab dataset, $|\mathbb{A}| = 6$, different $|\mathbb{M}|$}  
		\label{fig:VizPut-Cell_a_less_m}
	\end{subfigure}
	\begin{subfigure}{0.32\textwidth}
		\includegraphics[width=\linewidth]{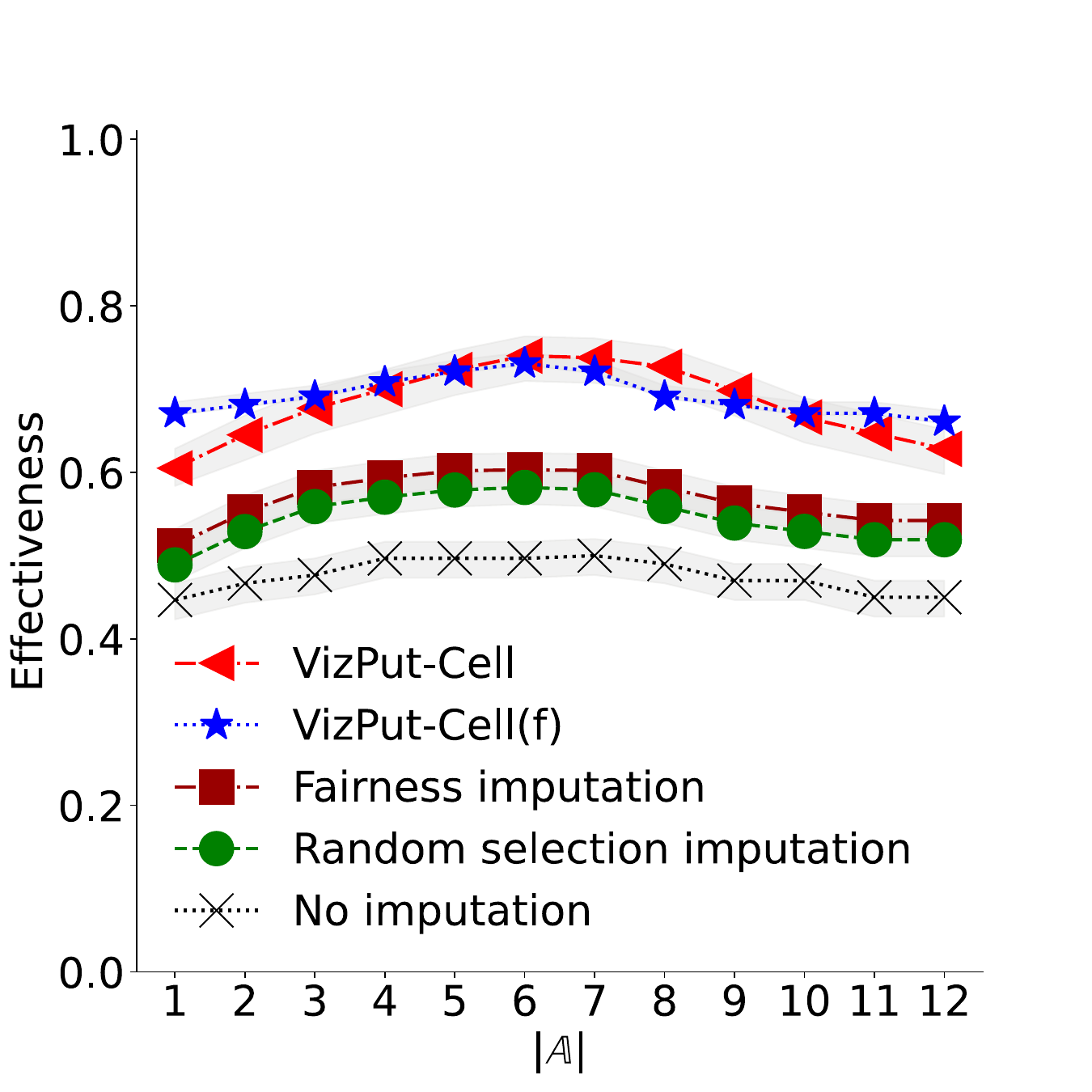}
		\caption{Diab dataset, $|\mathbb{M}| = 6$, different $|\mathbb{A}|$}  
		\label{fig:VizPut-Cell_extrem}
	\end{subfigure}
	\caption{\textit{VizPut-Cell} vs \textit{VizPut-Cell(f)}, distance metric: RBO, with different settings of $|\mathbb{A}|$ and $|\mathbb{M}|$, and different datasets} 
	\label{fig:VizPut-Cell_size_a_m}
\end{figure*}

\noindent{\textbf{Effect of Fairness on Proposed Algorithms}}. Figure \ref{fig:VizPut-Cell_size_a_m} evaluates the performance of our proposed algorithms with and without fairness, considering various sizes of $|\mathbb{A}|$ and $|\mathbb{M}|$ as well as datasets. The experimental results reveal the following observations: 1) \textit{VizPut-Cell} surpasses \textit{VizPut-Cell(f)} when the sizes of $|\mathbb{A}|$ and $|\mathbb{M}|$ are identical; 2) under extreme conditions, such as when $|\mathbb{A}|$ is small and $|\mathbb{M}|$ is large, \textit{VizPut-Cell(f)} exhibits superior performance compared to \textit{VizPut-Cell}. The rationale behind these findings is that \textit{VizPut-Cell} selects missing cells based on their maximum contribution to the recommendation results (as described in Sec.~\ref{sec:insight_aware_approach}), potentially leading to an imbalance in imputation under extreme conditions. Moreover, \textit{VizPut-Cell(f)} demonstrates consistent performance across all conditions, including extreme ones. Therefore, it is advisable to employ \textit{VizPut-Cell(f)}, particularly when the number of dimensional and measure attributes are unequal.

\begin{figure*}[t!]
\centering
	\begin{subfigure}{0.32\textwidth}
		\includegraphics[width=\linewidth]{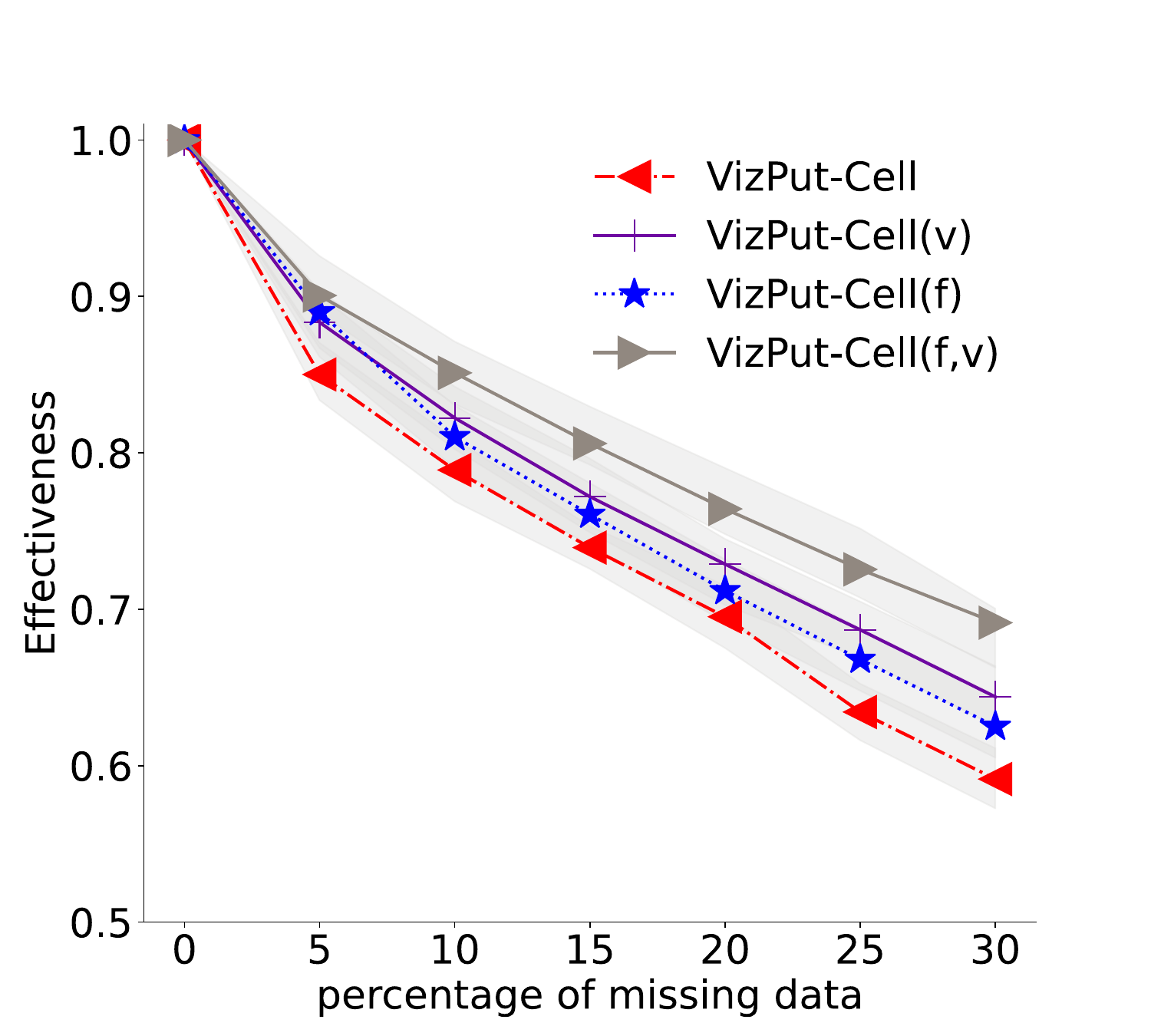}
		\caption{Case 1: $D_Q \subset D_R $} 
		\label{fig:case1}
	\end{subfigure}
	\begin{subfigure}{0.32\textwidth}
		\includegraphics[width=\linewidth]{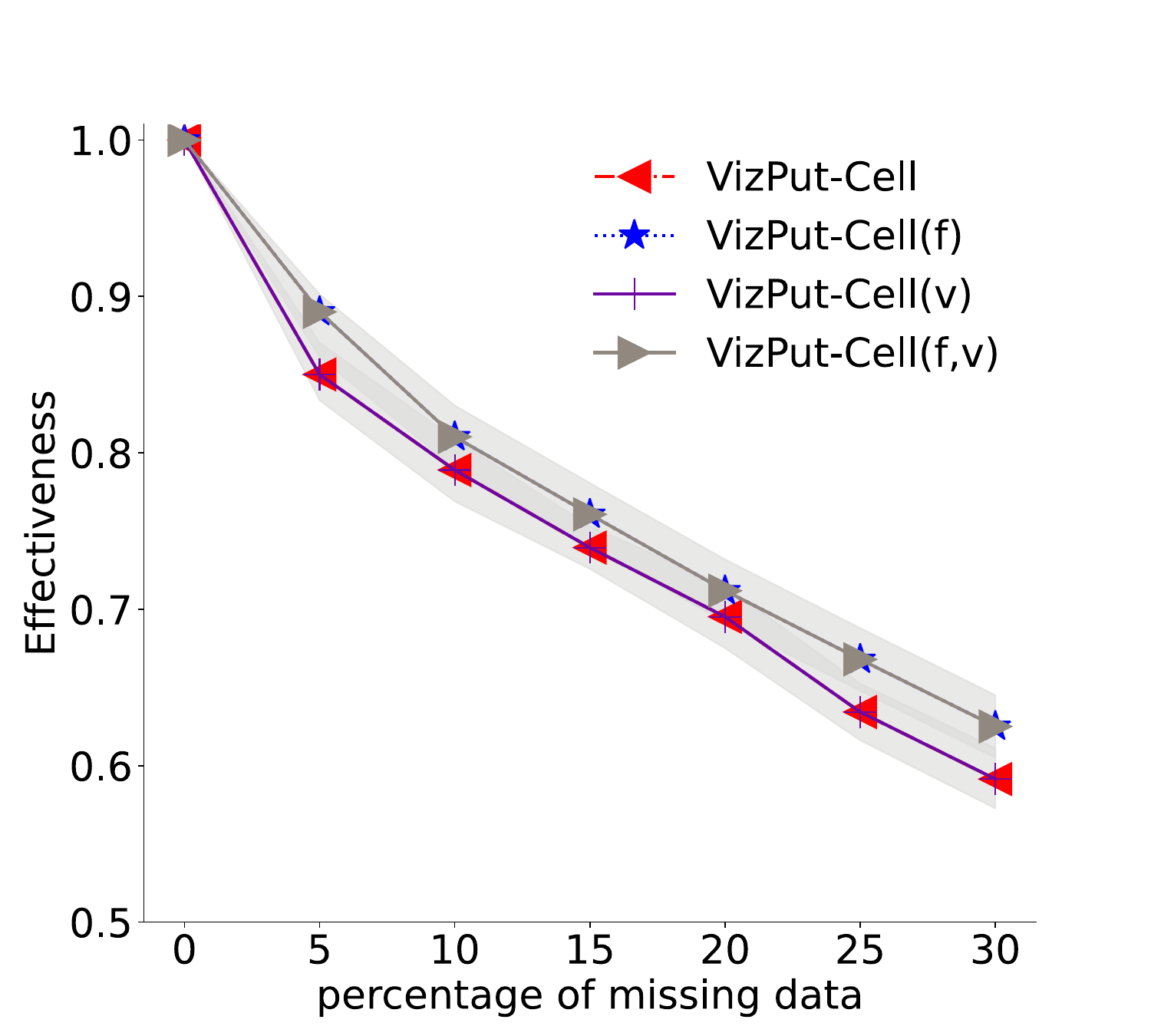}
		\caption{Case 2:$D_Q \cup D_R = D$}  
		\label{fig:case2}
	\end{subfigure}
	\begin{subfigure}{0.32\textwidth}
		\includegraphics[width=\linewidth]{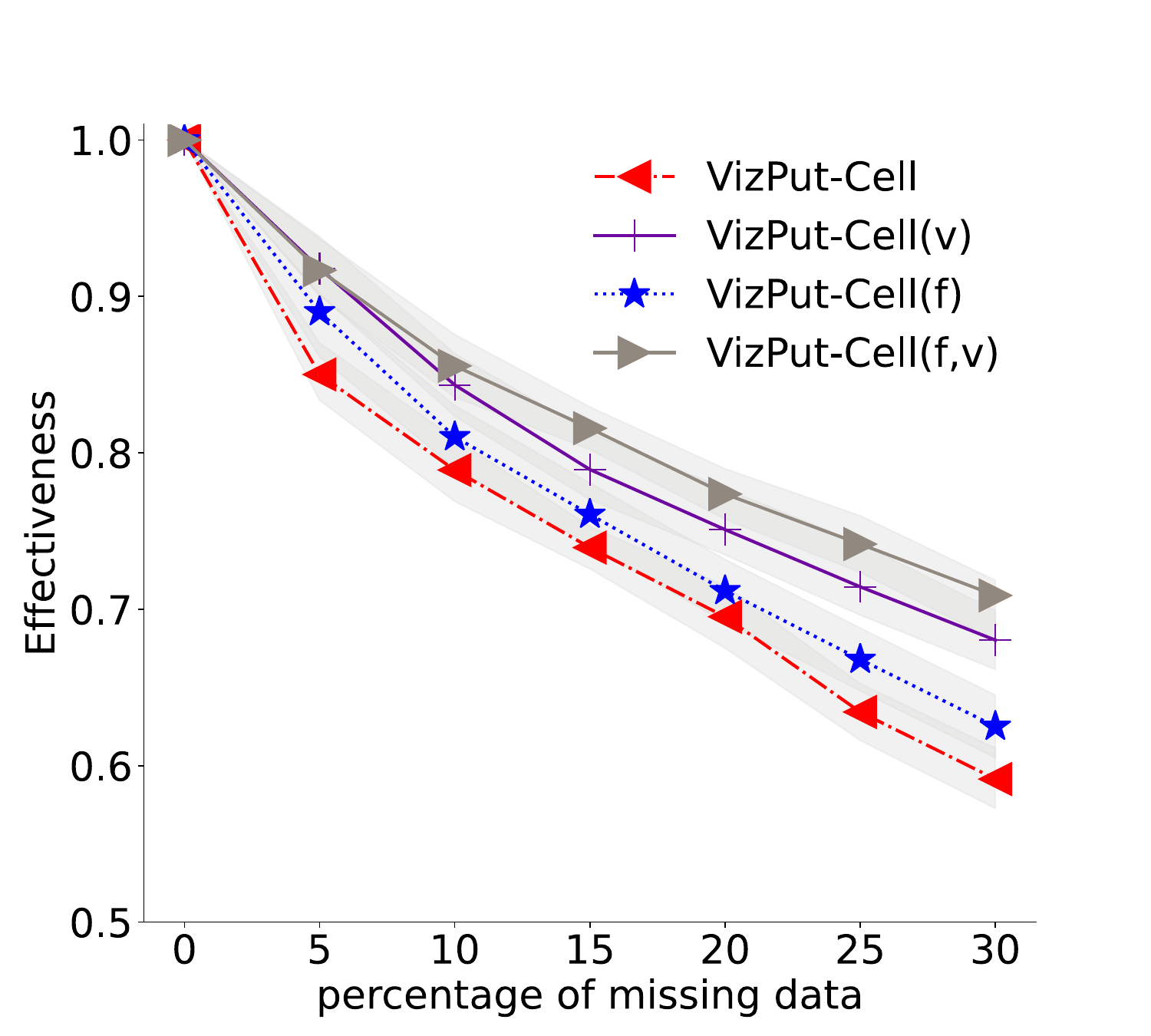}
		\caption{Case 3:$D_Q \cup D_R \neq D$}  
		\label{fig:case3}
	\end{subfigure}
	\caption{The impact of UsedView parameter $V$} 
	\label{fig:impact_used_view}
\end{figure*}

\noindent{\textbf{Impact of UsedView}}. In this experiment, we compare four algorithms: \textit{VizPut-Cell}, \textit{VizPut-Cell(f)}, \textit{VizPut-Cell(v)}, and \textit{VizPut-Cell(f, v)}. As discussed in Sec.~\ref{sec:insight_aware_approach}, there are three prevalent cases of target and reference subset settings in the top-$k$ visual insight recommendation task. The first case is $D_Q \subset D_R$, the second case is $D_Q \cup D_R = D$, and the third case is $D_Q \cup D_R \neq D$. Figure \ref{fig:impact_used_view} illustrates the influence of $V$ on these three cases. In the second case (Figure \ref{fig:case2}), the performance of both \textit{VizPut-Cell} and \textit{VizPut-Cell(f)} remains unaltered when utilizing $V$, since $D_Q \cup D_R = D$ and the $disease$ attribute has only two categories ($[Yes, No]$), resulting in merely two possible subsets based on the $disease$ predicate. Consequently, the $V$ score for all missing cells on $D_I$ (i.e., $D_Q \cup D_R$) is equal to 1. In contrast, in both the first and third cases, incorporating $V$ into the priority function enhances performance, as evidenced by \textit{VizPut-Cell(v)} outperforming \textit{VizPut-Cell}, and \textit{VizPut-Cell(f, v)} surpassing \textit{VizPut-Cell(f)}.

\begin{figure*}[t!]
\centering
	\begin{subfigure}{0.45\textwidth}
		\includegraphics[width=\linewidth]{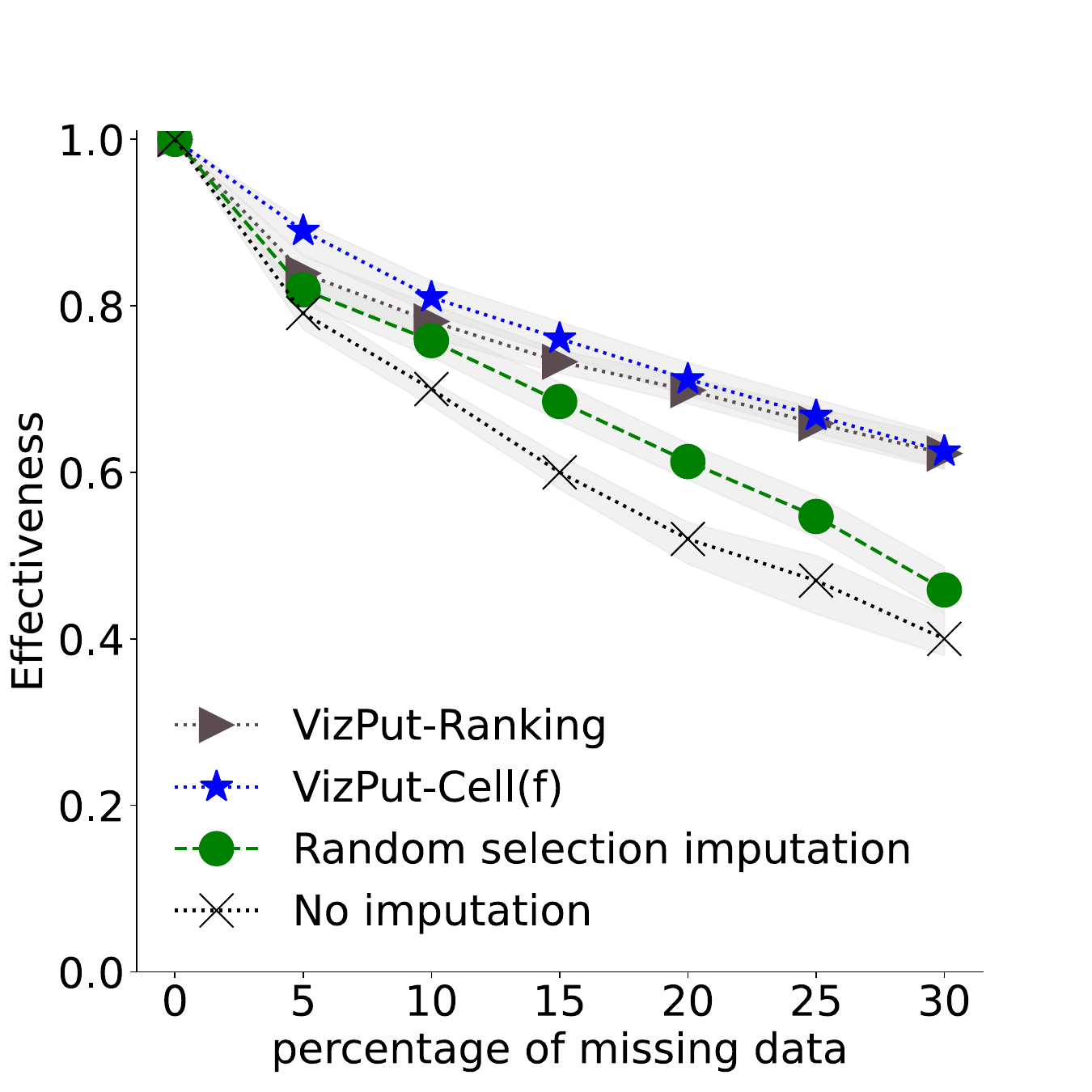}
		\caption{VizPut-Ranking vs. VizPut-Cell(f)} 
		\label{fig:VizPut-Ranking}
	\end{subfigure}
	\begin{subfigure}{0.45\textwidth}
		\includegraphics[width=\linewidth]{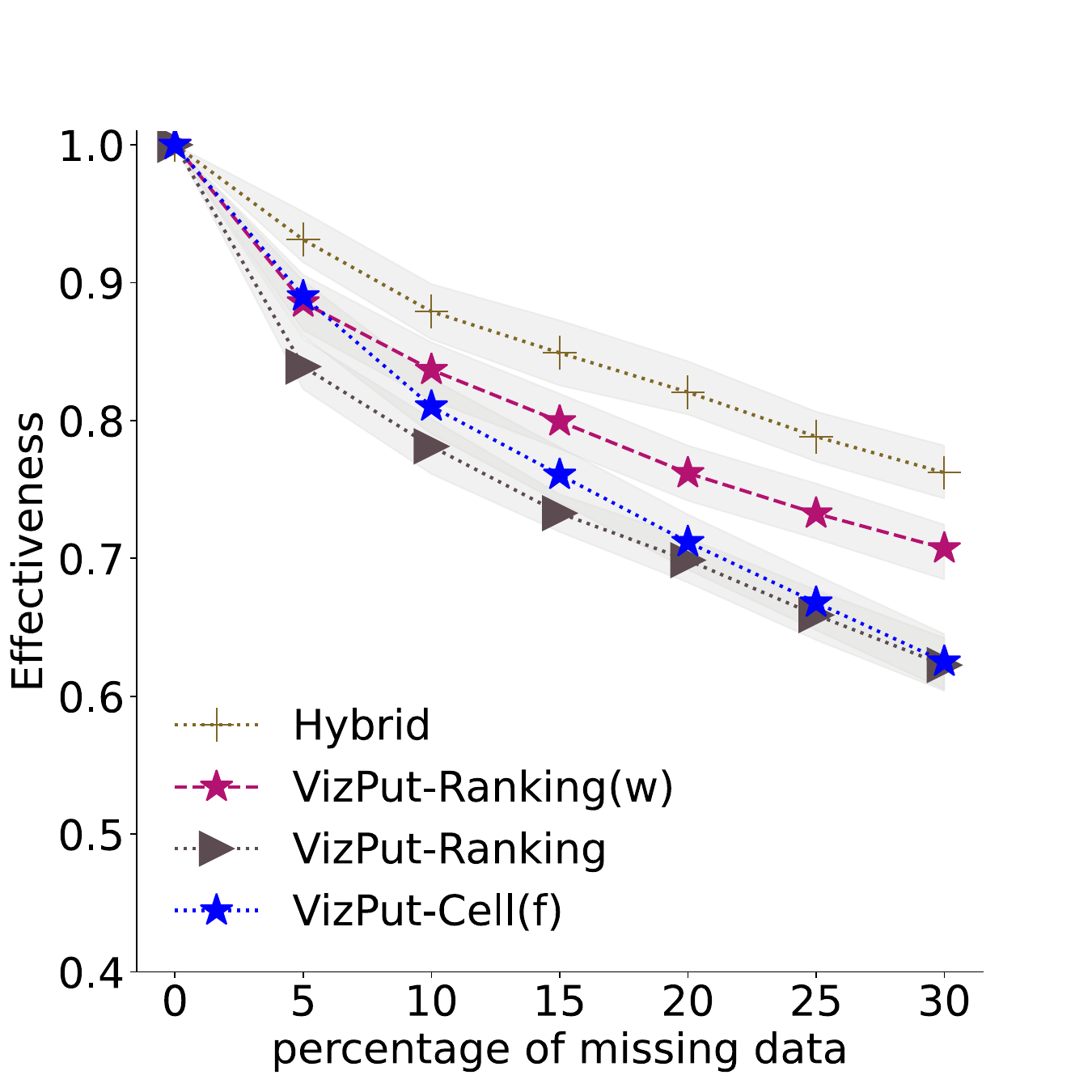}
		\caption{Hybrid vs standalone algorithms}  
		\label{fig:hybrid}
	\end{subfigure}
		\caption{The effectiveness of \textit{VizPut-Ranking, VizPut-Ranking(w), VizPut-Cell(f)} and \textit{Hybrid} approaches on heart disease dataset, where Hybrid is the combination of VizPut-Ranking(w) and VizPut-Cell(f)).}
	\label{fig:VizPut-Ranking_hybrid}
\end{figure*}

\noindent\textbf{Evaluating the Effectiveness of VizPut-Ranking and Hybrid Approaches}. Figure \ref{fig:VizPut-Ranking} presents the effectiveness of our proposed method, VizPut-Ranking, in comparison to VizPut-Cell(f) and the baselines (i.e., No Imputation, Random Selection Imputation, and Fairness Imputation), employing the RBO metric. The figure indicates that VizPut-Ranking surpasses the baselines; nevertheless, it is inferior to VizPut-Cell(f) when the percentage of missing values is low, and both exhibit similar performance when the percentage of missing values is around 30\%. Figure \ref{fig:hybrid} displays the effectiveness of VizPut-Ranking(w), revealing that this extended version outperforms the original VizPut-Ranking. This suggests that integrating a weight parameter can enhance the effectiveness of recommended visualizations. Figure \ref{fig:hybrid} also displays the performance of the hybrid approach, which combines VizPut-Ranking(w) and VizPut-Cell(f). Our hybrid approach outperforms the standalone algorithms, demonstrating that combining VizPut-Ranking(w) and VizPut-Cell(f) optimizes the handling of incomplete data for visualization recommendation.

\eat{
\begin{figure*}[t!]
\centering
	\begin{subfigure}{0.32\textwidth}
		\includegraphics[width=\linewidth]{newfigures/heart_disease_gap.pdf}
		\caption{Importance score gap of Heart disease dataset with different missing percentage} 
		\label{fig:gap_heart}
	\end{subfigure}
	\begin{subfigure}{0.45\textwidth}
		\includegraphics[width=\linewidth]{newfigures/heart_disease_avg_gap.pdf}
		\caption{The average of importance score gap of heart disease dataset}  
		\label{fig:avg_i_gap_heart}
	\end{subfigure}
	\begin{subfigure}{0.45\textwidth}
		\includegraphics[width=\linewidth]{newfigures/diab_avg_gap.pdf}
		\caption{The average of importance score gap of diabetes disease dataset}  
		\label{fig:avg_i_gap_diab}
	\end{subfigure}
		\caption{The importance score gap of the recommended visualizations generated from heart disease and diabetes dataset}
	\label{fig:i_gap}
\end{figure*}

\noindent\textbf{The importance score gap of the recommended visualizations generated from heart disease and diabetes dataset}. Figure \ref{fig:i_gap} presents the importance score gap of the recommended visualizations generated from heart disease and diabetes dataset.

}

\begin{figure*}[t!]
\centering
	\begin{subfigure}{0.45\textwidth}
		\includegraphics[width=\linewidth]{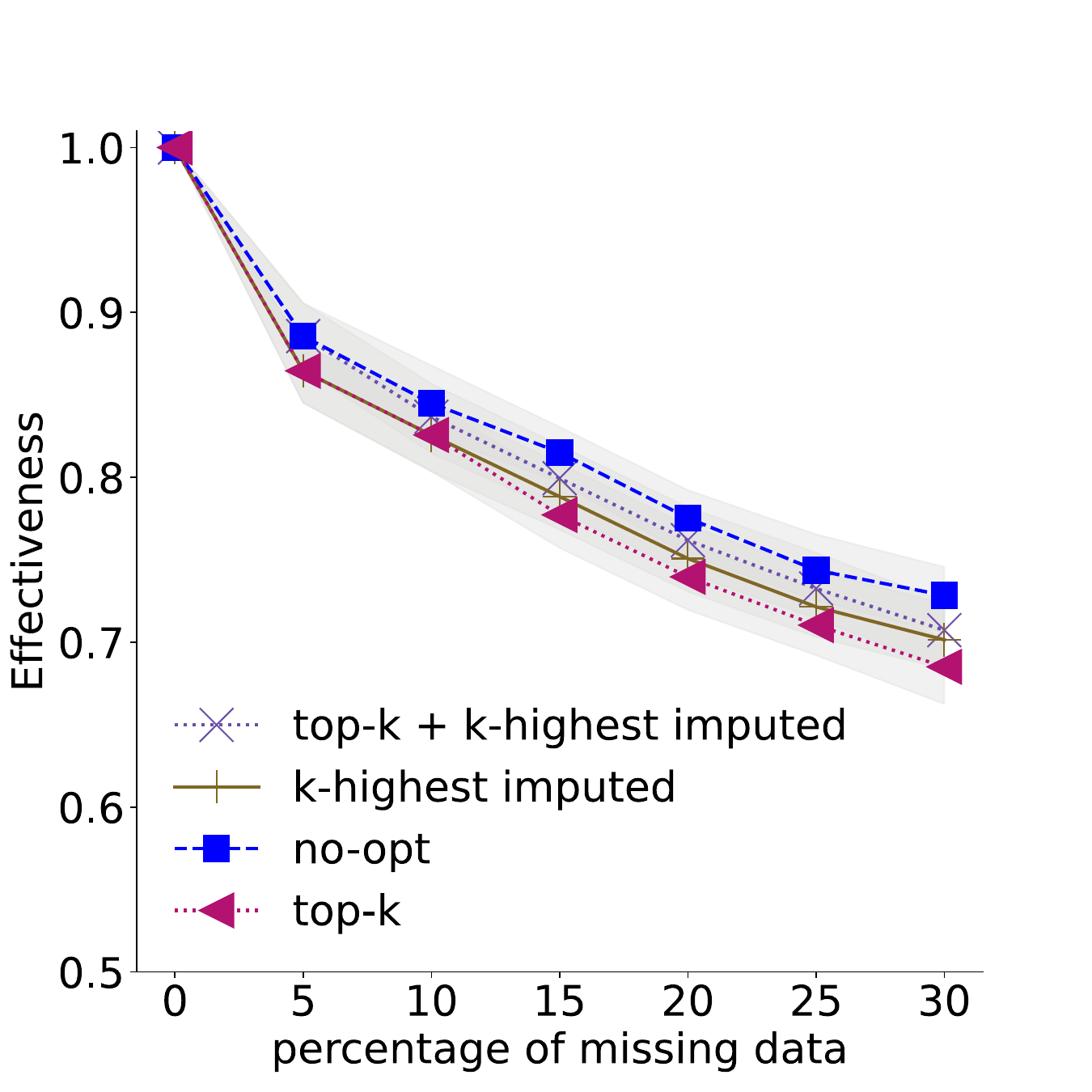}
		\caption{The effectiveness of VizPut-Ranking(w) with different optimization techniques for generating the final recommendation} 
		\label{fig:VizPut_Ranking_w_vness}
	\end{subfigure}
	\begin{subfigure}{0.45\textwidth}
		\includegraphics[width=\linewidth]{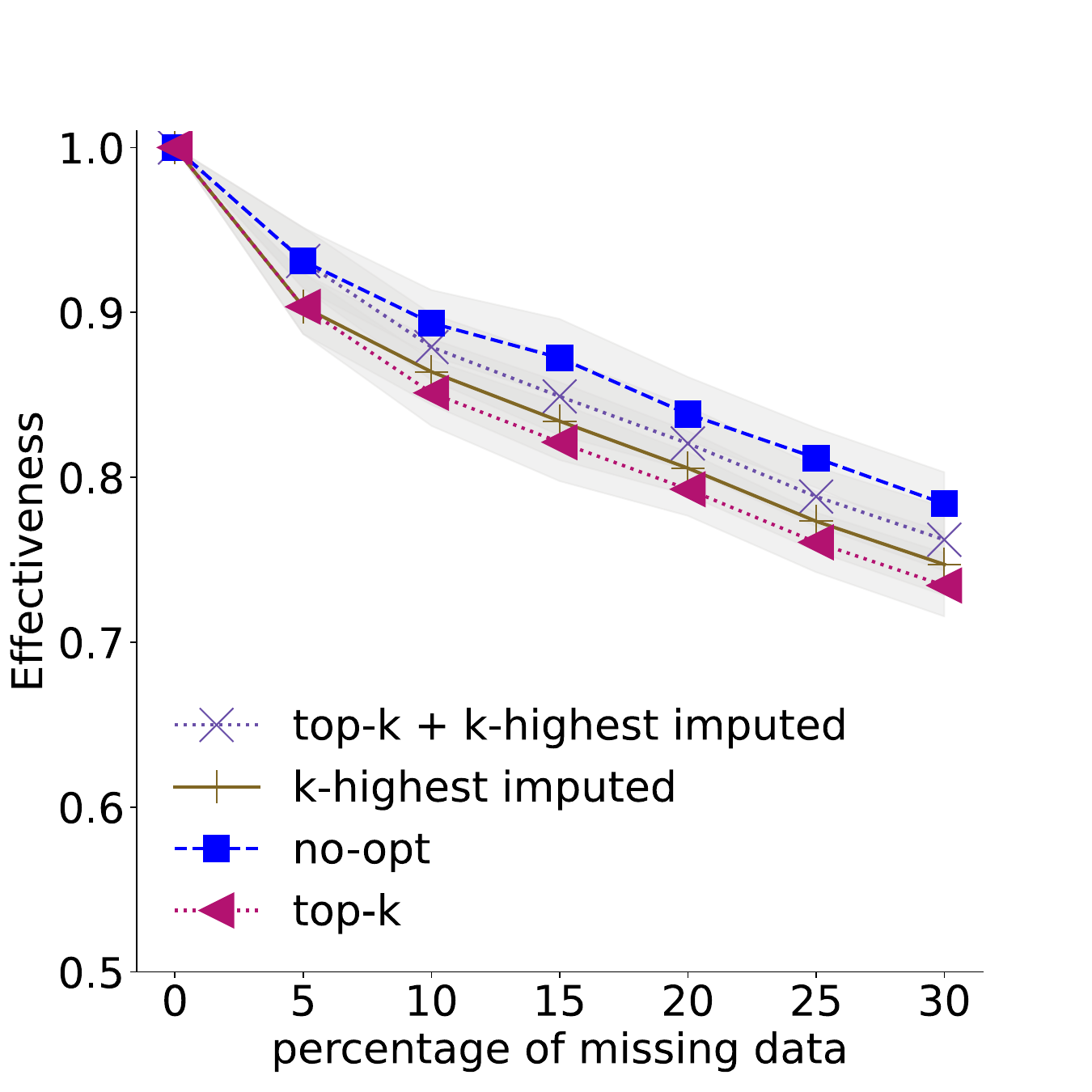}
		\caption{The effectiveness of Hybrid with different optimization techniques for generating the final recommendation}  
		\label{fig:VizPut_hybrid_vness}
	\end{subfigure}
		\caption{The effectiveness comparison of different optimization techniques for generating the final recommendation} 
	\label{fig:VizPut-insight-secondly-effectiveness}
\end{figure*}

\begin{figure*}[t!]
\centering
	\begin{subfigure}{0.45\textwidth}
		\includegraphics[width=\linewidth]{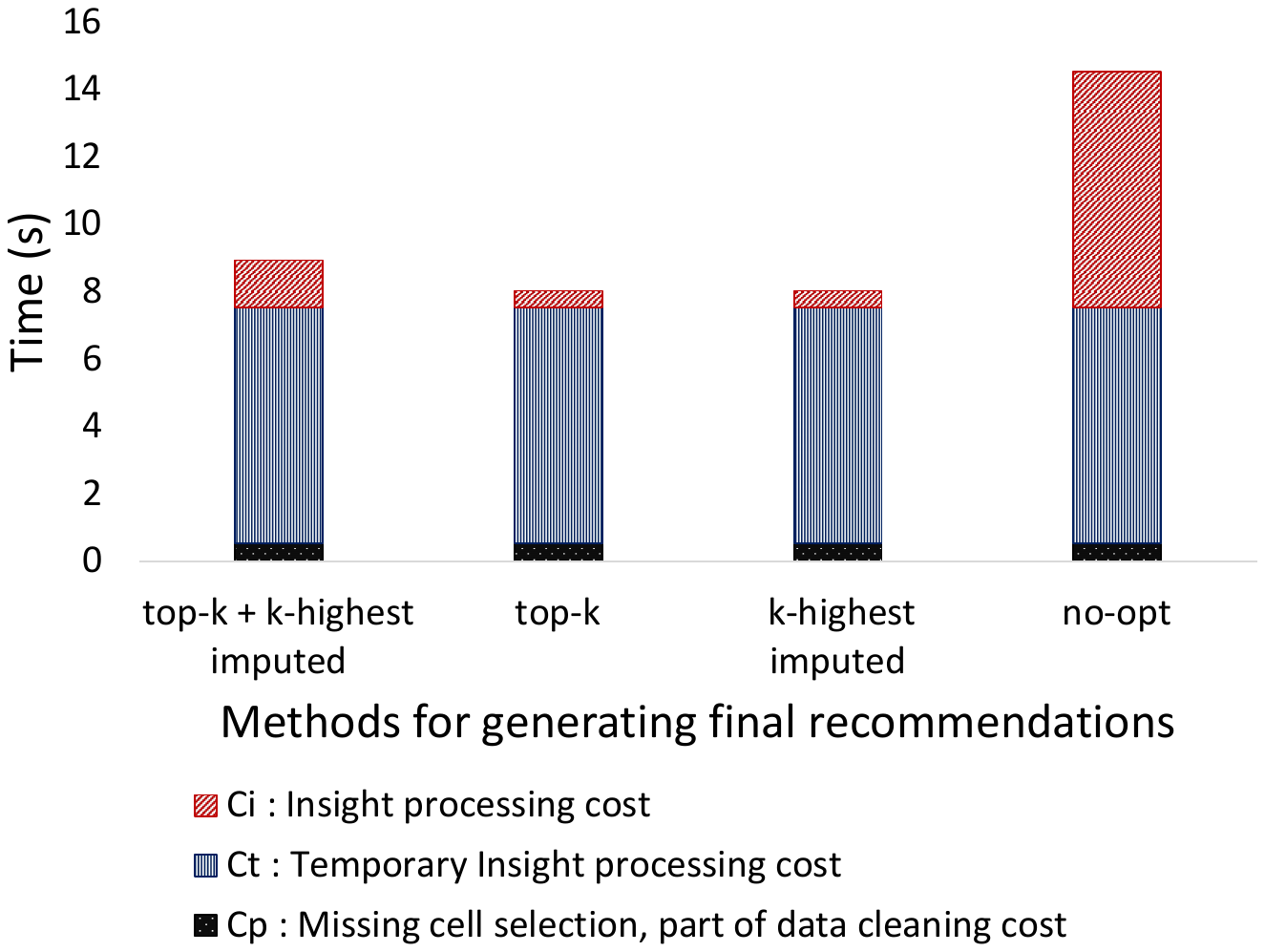}
		\caption{The cost of VizPut-Ranking(w)} 
		\label{fig:VizPut_Ranking_w}
	\end{subfigure}
	\begin{subfigure}{0.45\textwidth}
		\includegraphics[width=\linewidth]{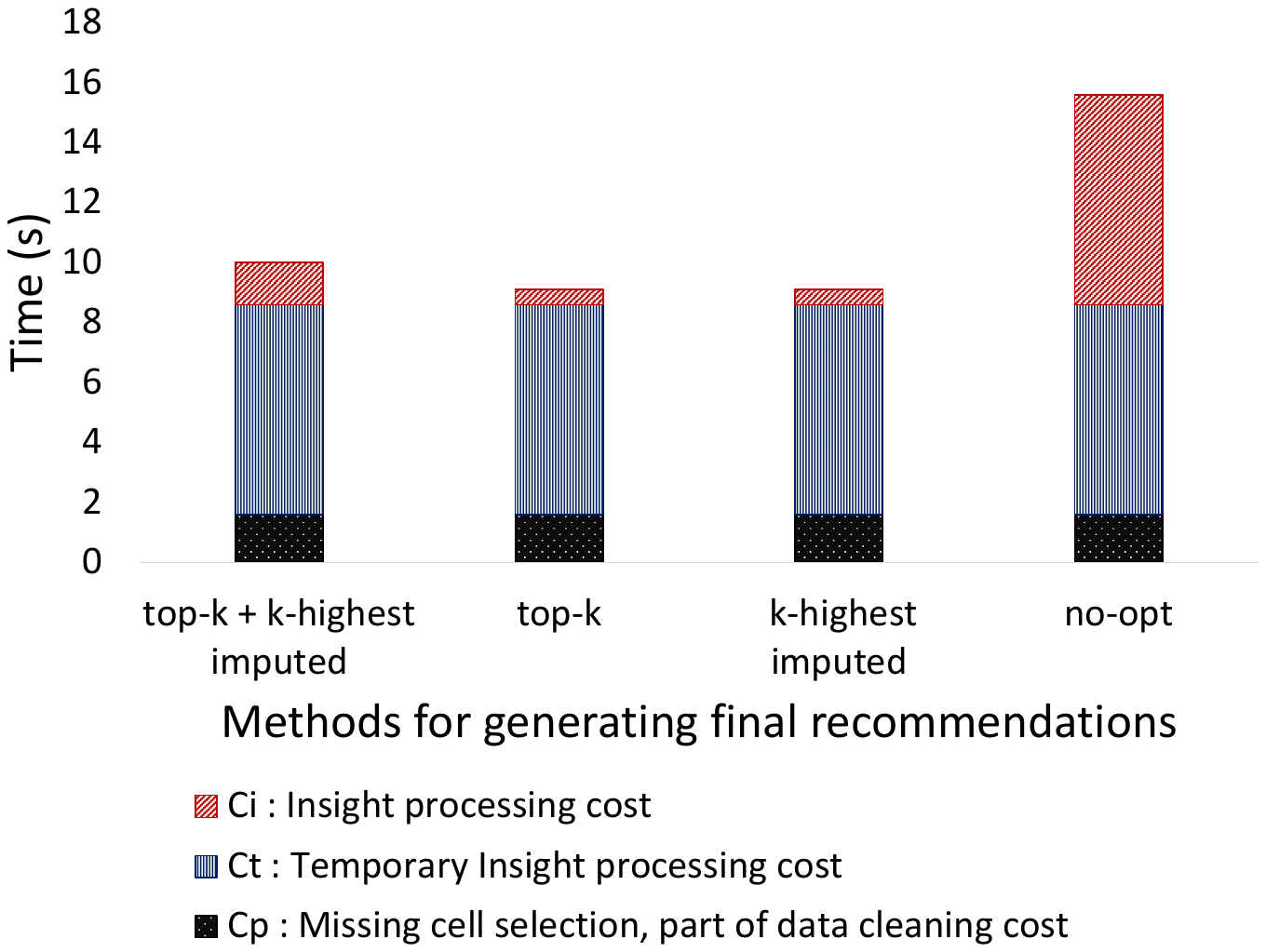}
		\caption{The cost of Hybrid}  
		\label{fig:VizPut_hybrid}
	\end{subfigure}
		\caption{Comparison of total cost between VizPut-Ranking(w) and the Hybrid method under different settings for the second insight generation from Airbnb dataset.} 
	\label{fig:VizPut-insight-secondly}
\end{figure*}

\noindent{\textbf{Effectiveness Comparison of Methods to Minimize Insight Processing Computation in VizPut-Ranking(w) and Hybrid}}. Figure \ref{fig:VizPut-insight-secondly-effectiveness} shows the effectiveness comparison of different optimization techniques for minimizing insight processing in both VizPut-Ranking(w) and Hybrid. Both figures indicate that the \textit{no-opt} approach, which requires the regeneration of all visualizations in the second generation (i.e., after imputation), has the highest effectiveness compared to the other proposed methods. However, this method is costly (see Figure \ref{fig:VizPut-insight-secondly}) due to its double insight processing expense. Therefore, we introduce optimization techniques that offer effectiveness close to \textit{no-opt} but at a more reasonable cost. Additionally, our proposed method—regenerating the top-k from temp-topk and the visualizations with the highest imputation (i.e.,\textit{ top-k + k-highest imputed})—exhibits superior effectiveness compared to other methods, with the exception of \textit{no-opt}. When comparing the costs of \textit{top-k + k-highest imputed} and \textit{no-opt} in Figure \ref{fig:VizPut-insight-secondly}, it is evident that \textit{top-k + k-highest imputed} is significantly more affordable than \textit{no-opt}. This cost difference is further elaborated upon in the subsequent section.

\noindent{\textbf{Cost Comparison of Methods to Minimize Insight Processing Computation in VizPut-Ranking(w) and Hybrid}}. As illustrated in Figure \ref{fig:VizPut-insight-secondly}, \textit{no-opt} requires twice the computation for insight generation processing. In this work, we rely on the \textit{top-k + k-highest imputed} method, which can significantly reduce the \(C_i\) value in comparison to the \textit{no-opt}. Note that we can also utilize the \textit{top-k} method, which only regenerates the top $k$ visualizations from \textit{temp-rank}, or the \textit{k-highest imputed} method, which only regenerates the top $k$ visualizations that received the highest imputation. While both of these approaches are more efficient than \textit{top-k + k-highest imputed}, they exhibit lower effectiveness. For instance, consider the Heart Disease dataset. For each single query input, it generates \(2 \times 8 \times 6 \times 4 = 384\) visualizations. The baseline solution (\textit{no-opt}) would re-execute all 384 visualizations. On the other hand, the \textit{top-k} method only selects \(k=10\) visualizations to regenerate, similar to the \textit{k-highest imputed} method, which also regenerates \(k=10\) visualizations. It is also noted that some visualizations from \textit{top-k + k-highest imputed} often intersect; thus, the combination of \textit{top-k + k-highest imputed} might regenerate at most 20 visualizations, but usually, it is less than that.
Considering the context of the Airbnb dataset, which has fewer attributes compared to the Heart Disease dataset, however, it has a higher number of tuples, it serves well for efficiency checking. Similar to the Heart Disease dataset, a single query input in the Airbnb dataset generates \(2 \times 4 \times 4 \times 4= 128\) visualizations. Using the \textit{no-opt} method would lead to the regeneration of all 128 visualizations. However, employing the \textit{top-k} or \textit{k-highest imputed} methods would mean only 10 visualizations are regenerated in each case. Combining \textit{top-k + k-highest imputed} could regenerate a maximum of 20 visualizations, though it often results in a number less than that.

\begin{figure*}[t!]
\centering
 \begin{subfigure}{0.45\textwidth}
		\includegraphics[width=\linewidth]{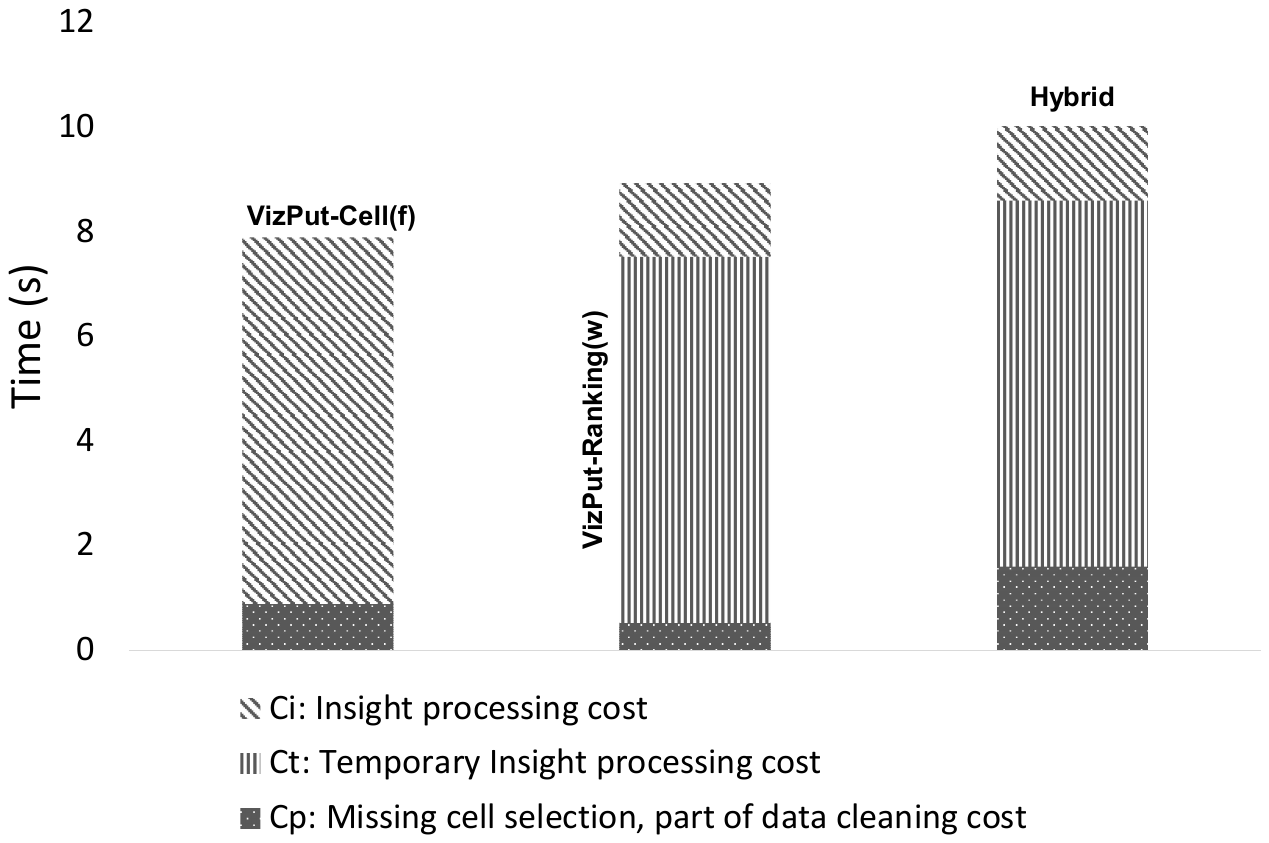}
		\caption{Airbnb dataset}  
		\label{fig:airbnb_dataset_cost}
	\end{subfigure}
	\begin{subfigure}{0.45\textwidth}
		\includegraphics[width=\linewidth]{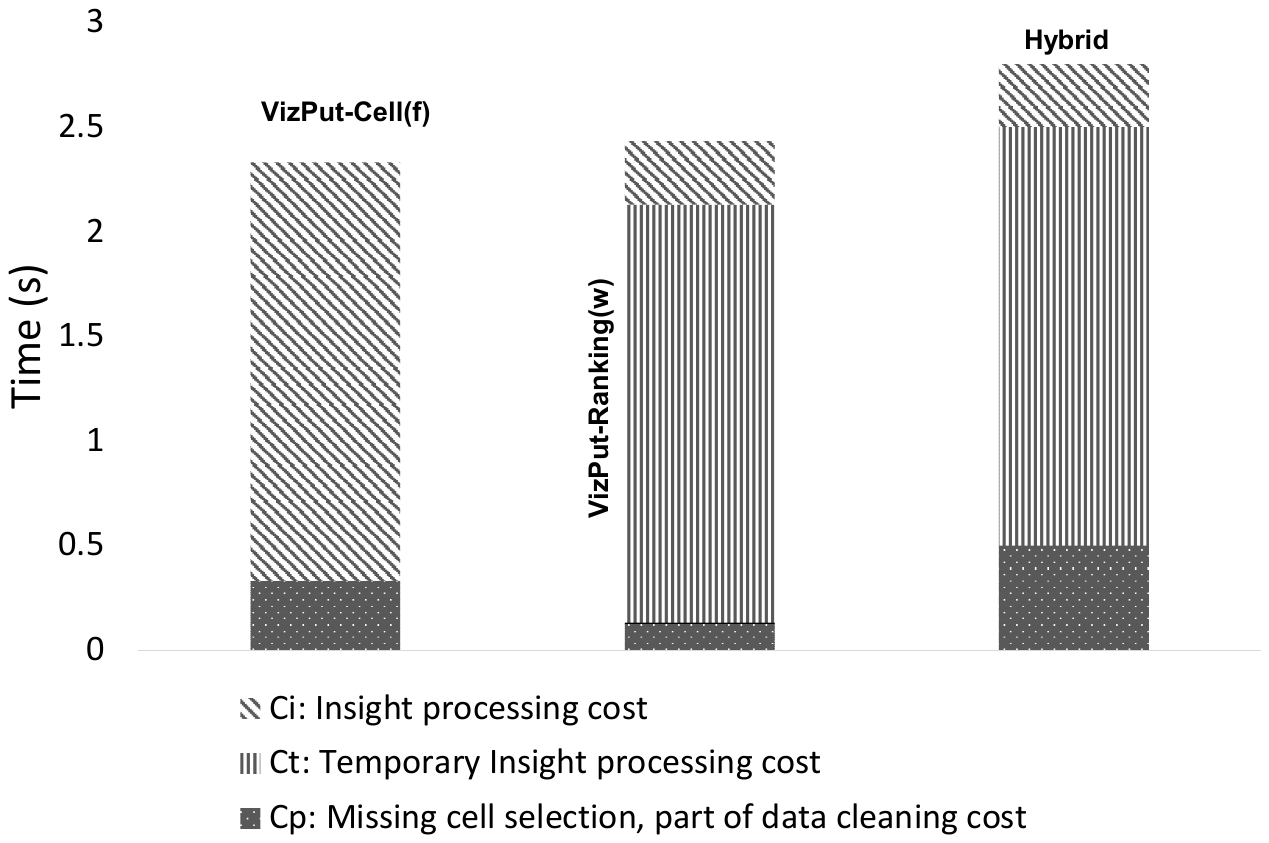}
		\caption{Heart disease dataset} 
		\label{fig:heart_dataset_cost}
	\end{subfigure}
		\caption{Total cost comparison of VizPut-Cell(f), VizPut-Ranking(w), and Hybrid with different datasets} 
	\label{fig:VizPut-Ranking_hybrid}
\end{figure*}

\noindent{\textbf{Cost Comparison of VizPut-Cell(f), VizPut-Ranking(w), and Hybrid}}. As depicted in Figure \ref{fig:hybrid}, both VizPut-Ranking(w) and Hybrid offer advantages in terms of effectiveness, with the Hybrid method exhibiting superior effectiveness compared to the other methods. However, considering efficiency, VizPut-Cell(f) clearly has the lowest cost compared to the other proposed methods, as demonstrated in Figures~\ref{fig:heart_dataset_cost} and~\ref{fig:airbnb_dataset_cost}. The efficiency of VizPut-Cell(f) can be attributed to its mechanism of generating recommendations only once, in contrast to VizPut-Ranking(w) and Hybrid, which generate recommendations twice—initially producing a temporary rank prior to data imputation, and then creating final recommendations. Although VizPut-Ranking(w) and Hybrid generate recommendations twice, our proposed optimization techniques reduce the cost associated with the second generation of insights. This cost reduction is achieved by regenerating only $k$ views, determined by the user-provided $k$ parameter and the $k$ views with the highest imputation budgets. Furthermore, among all the approaches, the Hybrid method incurs the highest cost for missing cell selection $C_p$ (a component of the data cleaning cost $C_c$). This increased cost arises because the Hybrid approach integrates multiple priority functions, each with its unique method to prioritize the imputation of missing cells.

\vspace{-10pt}
\section{Conclusions}
\vspace{-10pt}
This paper presents three methods for dealing with incomplete data in visualization recommendations. Users can choose a method depending on their needs and preferences. The \textit{Cell-aware VizPut} variants are suitable for users who prefer to impute data first and then generate recommendations ("\textit{impute-first-insight-next}" approach). On the other hand, the \textit{Ranking-aware VizPut} variants follow the opposite process, generating insights first, obtaining the temporary top-k \textit{temp-rank}, and imputing based on it. The hybrid approach combines both \textit{Cell-aware VizPut} and \textit{Ranking-aware VizPut} to achieve maximum performance in handling incomplete data in visualization recommendations.

\smallskip
{\bf Acknowledgments: } Rischan Mafrur is sponsored by the Indonesia Endowment Fund for Education (Lembaga Pengelola Dana Pendidikan/LPDP, scholarship ID: 201706220111044). A/Prof Mohamed A. Sharaf is supported by UAE University Grant (G00003352). Prof Guido Zuccon is the recipient of an Australian Research Council DECRA Research Fellowship (DE180101579) and a Google Faculty Award. 

\vspace{-10pt}
\bibliographystyle{splncs04}
\bibliography{insight_aware_paper}

\begin{thebibliography}{10}
\providecommand{\url}[1]{\texttt{#1}}
\providecommand{\urlprefix}{URL }
\providecommand{\doi}[1]{https://doi.org/#1}

\bibitem{Tableau}
Data in. brilliance out., \url{https://public.tableau.com/s/}

\bibitem{DiabDataset}
Diabetes 130 us hospitals 1999-2008, \url{https://www.kaggle.com/brandao/diabetes}

\bibitem{HeartDisease}
Heart disease data set, \url{https://archive.ics.uci.edu/ml/datasets/heart+Disease}

\bibitem{airbnb}
Inside airbnb, \url{http://insideairbnb.com/new-york-city/}

\bibitem{PowerBI}
Power bi | interactive data visualization bi tools, \url{https://powerbi.microsoft.com/en-us/}

\bibitem{Spotfire}
Spotfire: An information exploration environment., \url{https://www.tibco.com/products/tibco-spotfire/}

\bibitem{DynamicImputation}
Cambronero, J., Feser, J.K., Smith, M., Madden, S.: Query optimization for dynamic imputation. {PVLDB}  \textbf{10}(11),  1310--1321 (2017)

\bibitem{nadeef}
Dallachiesa, M., Ebaid, A., Eldawy, A., Elmagarmid, A.K., Ilyas, I.F., Ouzzani, M., Tang, N.: {NADEEF:} a commodity data cleaning system. In: SIGMOD (2013)

\bibitem{Foresight}
Demiralp, {\c{C}}., Haas, P.J., Parthasarathy, S., Pedapati, T.: Foresight: Recommending visual insights. {PVLDB}  \textbf{10}(12),  1937--1940 (2017)

\bibitem{QuickInsight}
Ding, R., et~al.: Quickinsights: Quick and automatic discovery of insights from multi-dimensional data. In: SIGMOD (2019)

\bibitem{MuVE2016}
Ehsan, H., Sharaf, M.A., Chrysanthis, P.K.: Muve: Efficient multi-objective view recommendation for visual data exploration. In: ICDE (2016)

\bibitem{TKDEHumaira}
Ehsan, H., Sharaf, M.A., Chrysanthis, P.K.: Efficient recommendation of aggregate data visualizations. TKDE  \textbf{30}(2),  263--277 (2018)

\bibitem{mean_knn_imputation}
Farhangfar, A., Kurgan, L.A., Dy, J.G.: Impact of imputation of missing values on classification error for discrete data. Pattern Recognit.  \textbf{41}(12),  3692--3705 (2008)

\bibitem{VizGRank}
Gao, Q., He, Z., Jing, Y., Zhang, K., Wang, X.S.: Vizgrank: {A} context-aware visualization recommendation method based on inherent relations between visualizations. In: {DASFAA} (2021)

\bibitem{AQP}
Garofalakis, M.N., Gibbons, P.B.: Approximate query processing: Taming the terabytes. In: {VLDB} (2001)

\bibitem{SpotLight}
Harris, C., Rossi, R.A., Malik, S., Hoffswell, J., Du, F., Lee, T.Y., Koh, E., Zhao, H.: Insight-centric visualization recommendation. CoRR  \textbf{abs/2103.11297} (2021)

\bibitem{Profiler}
Kandel, S., Parikh, R., Paepcke, A., Hellerstein, J.M., Heer, J.: Profiler: integrated statistical analysis and visualization for data quality assessment. In: AVI (2012)

\bibitem{VizDeck}
Key, A., Howe, B., Perry, D., Aragon, C.R.: Vizdeck: self-organizing dashboards for visual analytics. In: SIGMOD (2012)

\bibitem{taxonomy_dirty_data}
Kim, W.Y., et~al.: A taxonomy of dirty data. KDD  \textbf{7}(1),  81--99 (2003)

\bibitem{clustering_imputation}
Liao, Z., Lu, X., Yang, T., Wang, H.: Missing data imputation: {A} fuzzy k-means clustering algorithm over sliding window. In: {FSKD} (2009)

\bibitem{statistical_imputation}
Little, R.J.A., et~al.: Statistical Analysis with Missing Data. John Wiley, USA (1986)

\bibitem{VisClean}
Luo, Y., Chai, C., Qin, X., Tang, N., Li, G.: Interactive cleaning for progressive visualization through composite questions. In: {ICDE} (2020)

\bibitem{DeepEye}
Luo, Y., Qin, X., Tang, N., Li, G.: Deepeye: Towards automatic data visualization. In: ICDE (2018)

\bibitem{DiVE}
Mafrur, R., Sharaf, M.A., Khan, H.A.: Dive: Diversifying view recommendation for visual data exploration. In: CIKM (2018)

\bibitem{QualityMatters}
Mafrur, R., Sharaf, M.A., Zuccon, G.: Quality matters: Understanding the impact of incomplete data on visualization recommendation. In: DEXA (2020)

\bibitem{Jaccard}
Manning, C.D., et~al.: Introduction to information retrieval. Cambridge (2008)

\bibitem{VisGNN}
Ojo, F., Rossi, R.A., Hoffswell, J., Guo, S., Du, F., Kim, S., Xiao, C., Koh, E.: Visgnn: Personalized visualization recommendation via graph neural networks. In: {WWW}. pp. 2810--2818. {ACM} (2022)

\bibitem{ML_partial_complete_data}
Razniewski, S., Korn, F., Nutt, W., Srivastava, D.: Identifying the extent of completeness of query answers over partially complete databases. In: {SIGMOD} (2015)

\bibitem{ML_imputation}
Silva{-}Ram{\'{\i}}rez, E., Pino{-}Mej{\'{\i}}as, R., L{\'{o}}pez{-}Coello, M., Cubiles{-}de{-}la{-}Vega, M.: Missing value imputation on missing completely at random data using multilayer perceptrons. Neural Networks  \textbf{24}(1),  121--129 (2011)

\bibitem{ExtractTopK}
Tang, B., Han, S., Yiu, M.L., Ding, R., Zhang, D.: Extracting top-k insights from multi-dimensional data. In: SIGMOD (2017)

\bibitem{SeeDB2014}
Vartak, M., Madden, S., Parameswaran, A.G., Polyzotis, N.: {SEEDB:} automatically generating query visualizations. In: {PVLDB} (2014)

\bibitem{SeeDB2015}
Vartak, M., Rahman, S., Madden, S., Parameswaran, A.G., Polyzotis, N.: {SEEDB:} efficient data-driven visualization recommendations to support visual analytics. In: {PVLDB} (2015)

\bibitem{SampleClean}
Wang, J., Krishnan, S., Franklin, M.J., Goldberg, K., Kraska, T., Milo, T.: A sample-and-clean framework for fast and accurate query processing on dirty data. In: {SIGMOD} (2014)

\bibitem{RBO}
Webber, W., et~al.: A similarity measure for indefinite rankings. TOIS  \textbf{28}(4),  20--38 (2010)

\bibitem{Viewseeker}
Zhang, X., Ge, X., Chrysanthis, P.K., Sharaf, M.A.: Viewseeker: An interactive view recommendation framework. Big Data Res.  \textbf{25},  100238 (2021)

\end{thebibliography}
\end{document}